\documentclass[aps,prr,amsmath,amssymb,longbibliography,twocolumn,floatfix]{revtex4-2}

\usepackage{graphicx}
\usepackage{dcolumn}
\usepackage{bm}
\usepackage{amsmath}
\usepackage{amssymb} 
\usepackage{physics}
\usepackage{color}
\usepackage[dvipsnames]{xcolor}
\usepackage{hyperref} \hypersetup{ colorlinks=true, linkcolor=blue}
\usepackage[normalem]{ulem}

\newcommand{\nbar}{\bar{n}}

\begin{document}

\title{The thermodynamic uncertainty relation of a quantum-mechanically coupled two-qubit system}

\author{Kwang Hyun Cho}%
\email[]{khcho@kias.re.kr}
\affiliation{Center for AI and Natural Sciences, Korea Institute for Advanced Study}
\author{Hyukjoon Kwon}%
\email[]{hjkwon@kias.re.kr}
\affiliation{School of Computational Sciences, Korea Institute for Advanced Study}
\author{Changbong Hyeon}%
\email[]{hyeoncb@kias.re.kr}
\affiliation{School of Computational Sciences, Korea Institute for Advanced Study}

\date{\today}
\begin{abstract}
The minimal bound of the thermodynamic uncertainty relation (TUR) is modulated from that of the classical counterpart ($\mathcal{Q}_{\rm min}=2$) when a quantumness is present in the dynamical process far from equilibrium. 
A recent study on a dissipative two-level system (TLS) subject to an external field indicates that quantum coherence can suppress the fluctuations of the irreversible current and loosens the TUR bound to $\mathcal{Q}_{\rm min}^{\rm TLS}\approx 1.25$.  
Here, we extend on the field-driven single TLS 
to a quantum-mechanically coupled two-qubit system (TQS), and explore how the quantum coupling between the two qubits{\color{black}, an additional complexity introduced to the probem of TLS,} affects the photon current,  fluctuations, and the TUR bound. 
We find that the TUR bound of TQS depends on the strength of coupling, such that $\mathcal{Q}_{\rm min}^{\rm TQS}=\mathcal{Q}_{\rm min}^{\rm TLS}\approx 1.25$ when the two qubits are effectively decoupled under weak coupling, whereas another loose bound $\mathcal{Q}_{\rm min}^{\rm TQS}\approx 1.36$ is identified for two strongly coupled qubits under  strong fields. 
By contrasting the TQS against two coupled noisy oscillators, we illuminate the quantumness unique to the TQS and its effect on the TUR. {\color{black}Our findings from the study of TQS form the basis for understanding the TUR of more general $N$-qubit systems.}  
\end{abstract}
\maketitle




\section{introduction}
Since its first proposal~\cite{barato2015PRL} and subsequent proof for continuous time Markov jump processes on networks and overdamped Langevin dynamics~\cite{Pigolotti2017PRL,Gingrich2016PRL}, 
the thermodynamic uncertainty relation (TUR) has been extended to more general contexts along with its variations over the past decade~\cite{pietzonka2016JSM,Proesmans:2017,Hyeon2017PRE,dechant2018PRE,horowitz2017proof,hasegawa2019PRL,hasegawa2019PRE,koyuk2019PRL,dechant2018multidimensional,gingrich2017PRL,Hwang2018JPCL,lee2018PRE,pal2021thermodynamic,Song2020JPCL,lee2021universal,Song2021JCP,song2021cost,kwon2022thermodynamic,ziyin2023universal}.  
{\color{black}The TURs are written in the form of an inequality of a product ($\mathcal{Q}$) between total entropy production, $\Delta S_{tot}(t)/k_B$, and the squared relative error of a current-like dynamical observable $\theta(t)$ with odd parity $\theta(-t)=-\theta(t)$, satisfying  
\begin{align}
\mathcal{Q}= \frac{\Delta S_{tot}(t)}{k
_B}\frac{\langle\delta\theta^2(t)\rangle}{\langle\theta(t)\rangle^2}\geq \mathcal{Q}_{\rm min}. 
\label{eqn:general_TUR}
\end{align}
This signifies the trade-off between the dissipation and the precision of a dynamical process, with their uncertainty product $\mathcal{Q}$ being lower-bounded.  
If one were to increase the precision of the process, one should obtain a longer time trace, which requires greater amount of energetic cost. 
The relation (Eq.~\eqref{eqn:general_TUR}) can also be interpreted as a quantitative version of thermodynamic principle for nonequilibrium processes that constrains the total entropy production more tightly than the second law as $\Delta S_{tot}(t)/k
_B\geq \mathcal{Q}_{\rm min}\langle\theta(t)\rangle^2/\langle\delta\theta^2(t)\rangle$~\cite{dechant2018PRE,horowitz2019NaturePhys}. 
The minimal bound is $\mathcal{Q}_{\rm min}=2$, for Markov jump process and overdamped Langevin systems under constant driving, which encompass a number of conventional situations in nonequilibrium steady states; however, the value of $\mathcal{Q}_{\rm min}$ further reduces for more general nonequilibrium processes, when the system of interest is underdamped~\cite{kwon2022thermodynamic,fischer2020PRE,lee2019PRE} or subject to time-dependent drivings~\cite{koyuk2019PRL,koyuk2020PRL}.     
}

Several studies that examined the TURs for open quantum processes have indicated a loosening of its minimal bound {\color{black}$\mathcal{Q}_{\rm min}<2$}, leading to the TUR violation~\cite{agarwalla2018PRB,liu2019thermodynamic,kalaee2021PRE,hasegawa2021thermodynamic,ehrlich2021broadband,taddei2023thermodynamic,razzoli2024synchronization,mohan2025coherent,ohnmacht2025thermodynamic},  
signifying that quantumness can 
enhance the precision of the dynamical processes beyond its classical limit set by the conventional TUR.    
{\color{black}Specifically, studies of molecular junctions, made of a single or two quantum dots between the baths of thermal or chemical gradient driving heat or charge transport, examined the TUR or the upper-bound of engine efficiency, induced by the quantum coherence or particle correlation~\cite{brandner2018PRL,agarwalla2018PRB,liu2019thermodynamic,cangemi2020violation,kalaee2021PRE,um2022coherence,kirchberg2022JCP,prech2023entanglement}.}

Our recent works of a field-driven two-level system (TLS) and three-level $\Lambda$-system immersed in photonic bath~\cite{singh2021,singh2023fundamental}, {\color{black}focusing on the effect of quantumness accompanied with light-matter interaction on the TUR,} have shown that the size of the fluctuations in photon current is decided by the competition between the real and imaginary parts of the quantum coherence{\color{black}, corresponding to the off-diagonal elements of the density matrix,} that arises from a superposition between different states in a given basis~\cite{streltsov2017colloquium}.  
The imaginary part of the coherence, which was shown to be responsible for the generation of non-equilibrium current {\color{black}in open quantum systems}~\cite{wu2012efficient,yang2020steady,roden2016probability}, is linked to the dissipative flow of energy or absorption between states that  suppresses the current fluctuations. 
In contrast, the real part, related to the dephasing mechanism accompanied by population mixing,
amplifies the fluctuations~\cite{roden2016probability,singh2021,singh2023fundamental}. 

Here, {\color{black}we consider a straightforward extension of the one-qubit TLS, which could be related to a variety of physical platforms involving coherent driving of multipartite quantum systems, including not only the above-mentioned molecular junctions~\cite{brandner2018PRL,agarwalla2018PRB,liu2019thermodynamic,cangemi2020violation,kalaee2021PRE,um2022coherence,kirchberg2022JCP}, but also cold atoms in optical lattices~\cite{maschler2008ultracold},  circuit quantum electrodynamics systems~\cite{reiter2013steady,shah2024stabilizing}, and quantum dots in cavity~\cite{mitra2010entanglement,gu2023probing}. Specifically,} 
we study a theoretical model of quantum-mechanically coupled two-qubit system (TQS), in which two atoms with ground and excited states~\cite{reichle2006experimental} are in a photonic bath while an external field continuously irradiates only one of the two atoms.  
The dynamics in the first atom can affect 
the quantum state of the second atom due to the coupling, or it simply emits and dissipates a photon into the bath, the process of which can be formulated with the corresponding Lindblad equations. 

To study the TUR of the TQS, we first review the TUR of the field-driven dissipative TLS in a photonic bath. 
We calculate the photon current flowing from the light source to the bath through the TQS. 
We study how the TQS responds to the varying strengths of external field and the coupling strength, by explicitly calculating the steady-state populations and coherences over varying extent of detuning. 
In comparison with the TLS,  
the expressions for photon current and its fluctuations 
is rather complicated. Nevertheless, we will show that each of the photon currents generated from the two qubits can still be expressed in terms of the intra-qubit coherences and the correlated coherence between the two qubits. 

We characterize non-equilibrium responses of the TQS 
by means of the uncertainty product of TUR ($\mathcal{Q}$), 
finding that the maximal precision in the photon current is attained when both the strengths of field and two-qubit coupling are moderate. Our study offers physical insights into this finding by discussing the details of how the coherence, correlation, and entanglement of TQS arise from the varying strength of external field and quantum coupling of the two qubits, and contribute to its TUR. 

\section{Theoretical model}
\subsection{Two-qubit system subject to an external field and its evolution equation}
We consider a system that two quantum-mechanically coupled qubits, labeled 1 and 2, with respective transition frequencies $\omega_1$ and $\omega_2$, are in a photonic bath and subject to an external field irradiating the 1st qubit. 
The Hamiltonian for the total system reads 
\begin{align}
    H=H_S+H_K+H_\textrm{ext}(t)+H_B+H_{SB}. 
\end{align}
The system Hamiltonian consisting of two qubits is given as 
\begin{align}
H_S&=\left(\hbar/2\right)\left(\omega_1\sigma^1_z+\omega_2 \sigma^2_z\right). 
\end{align}
Here, $\sigma^i$s ($i=1$, 2) are the Pauli operators of the 1st and 2nd qubit. 
The quantum-mechanical coupling between the two qubits, which potentially gives rise to an  entanglement, is modeled using 
\begin{align}
    H_K=\hbar K \left(\sigma^1_+\sigma^2_-+\sigma^1_-\sigma^2_+ \right),
    \label{eqn:coupling_H}
\end{align}
where $K$ is the coupling strength.
The external field with the frequency $\omega$ irradiates the 1st qubit, resulting in an interaction energy Hamiltonian, {\color{black}is modeled semi-classically as} 
\begin{align}
    H_\textrm{ext}(t)&=-\vec{d}\cdot\vec{E}(r,t)\nonumber\\
    &=-\hbar \Omega\left(e^{i\omega t}+e^{-i\omega t}\right)\left(\sigma^1_{+}+\sigma^1_{-}\right),
    \label{eq:H_ext}
\end{align}
where the dipole operator  $\vec{d}=\vec{d}_{10}\sigma^1_++\vec{d}_{01}\sigma^1_-$ acts on the first qubit and 
the electric field is given by $\vec{E}(r,t)\simeq \vec{\epsilon}e^{i\omega t}+\vec{\epsilon}^\ast e^{-i\omega t}$. 
{\color{black}In evaluating the effect of $H_{\rm ext}(t)$ on the density matrix, 
we consider the
dipole approximation that the driving field is nearly \emph{constant}
over the scale of our interest, such that $e^{ik\cdot r} \simeq 1$~\cite{scully1997quantum}.}
Thus, 
the 1st qubit display unitary oscillation between the ground and the excited states with the Rabi frequency 
$\Omega=\vec{d}_{10}\cdot\vec{\epsilon}/\hbar=\vec{d}_{01}\cdot\vec{\epsilon}^\ast/\hbar$. 
The Hamiltonian for the surrounding bath is 
\begin{align}
H_B=\sum_{\textbf{k},\xi}\hbar\omega_{\textbf{k}} b^\dagger_{\textbf{k},\xi} b_{\textbf{k},\xi},
\end{align}
where $b^\dagger_{\textbf{k},\xi}$ and $b_{\textbf{k},\xi}$ are the creation and annihilation operators for the bath degrees of freedom represented by simple harmonic oscillators with the angular frequency $\omega_k$, and the corresponding hamiltonian for the bath is summed over the wave vector ${\bf k}$ and polarization $\xi$. 
Finally, the Hamiltonian for the system-bath interaction is given by  
\begin{align}
H_{SB}&=\sum_{\textbf{k},\xi}\hbar\left(g^{1}_{\textbf{k},\xi} b_{\textbf{k},\xi} \sigma^1_+ +(g^1_{\textbf{k},\xi})^\ast b^\dagger_{\textbf{k},\xi}\sigma^1_-\right.\nonumber\\
&\left.+g^{2}_{\textbf{k},\xi}b_{\textbf{k},\xi} \sigma^2_+ +(g^2_{\textbf{k},\xi})^\ast b^\dagger_{\textbf{k},\xi}\sigma^2_-\right),
\end{align}
where $g^{1,2}_{\textbf{k},\xi}$ denotes the strength of system-bath coupling for each of the two qubits.

The Lindblad equation that casts the system-bath interaction in a specific form is obtained by tracing out the bath degrees of freedom and approximating the hierarchical equation under the weak-system bath coupling~\cite{manzano2020AIP}. 
The reduced density operator for the system $\rho(t)$ undergoes a dynamic evolution, obeying the following equation along with the Lindblad dissipator, $\mathcal{D}\left(\rho(t)\right)$, acting on both qubits
\begin{equation}
    \frac{d\rho(t)}{dt}=-\frac{i}{\hbar}\left[H_S + H_K  
    +H_\textrm{ext},\rho(t)\right]+\mathcal{D}\left(\rho(t)\right),
\end{equation}
where $\mathcal{D}\left(\rho(t)\right)=\mathcal{D}_1(\rho(t))+\mathcal{D}_2(\rho(t))$
with  
\begin{align}
    \mathcal{D}_i(\rho(t))&=\gamma_i \left(\nbar_i+1\right)\left(\sigma^i_-\rho(t)\sigma^i_+ -\frac{1}{2}\left\{\sigma^i_+\sigma^i_-,\rho(t)\right\}\right)\nonumber\\
    &+\gamma_i \nbar_i \left(\sigma^i_+\rho(t)\sigma^i_- -\frac{1}{2}\left\{\sigma^i_-\sigma^i_+,\rho(t)\right\}\right)
\end{align} 
for $i=1,2$. 
Here, the term $\bar{n}_i=(e^{\beta\hbar\omega_i}-1)^{-1}$ denotes the mean occupation number of the bosonic bath affected by the dynamics of the $i$-th qubit, 
$\{ A, B \}= A B + B A$
is the anti-commutator, 
and $\gamma_{1,2}$ is the relaxation rate of the qubit to its ground state caused by the interaction with the bath.  
The condition of the weak system-bath coupling is dictated by the condition of $\omega\gg\gamma_1,\gamma_2$. 
Rescaling the time scale by $\gamma_1$ as $t\rightarrow\tau=\gamma_1t$, and all the frequencies discussed here as $\gamma_2\rightarrow \gamma\equiv \gamma_2/\gamma_1$, $\Omega\rightarrow\Omega/\gamma_1$, {\color{black}and $K\rightarrow K/\gamma_1$} yield a set of coupled equations for the density matrix elements. 

Using 4 computational basis states $\ket{00}$, $\ket{01}$, $\ket{10}$, and $\ket{11}$, where $\ket{ab}\equiv \ket{a}_1\otimes\ket{b}_2$ signifies a tensor product that the 1st qubit is in the $a$ state and the 2nd qubit in the $b$ state, we express the state of TQS by the density matrix
\begin{align}
\rho &= \sum_{p,q,r,s} \rho_{pq,rs} \ket{pq}\bra{rs} \nonumber\\
&= \begin{pmatrix}
\rho_{11,11}&\rho_{11,10}&\rho_{11,01}&\rho_{11,00}\\
\rho_{10,11}&\rho_{10,10}&\rho_{10,01}&\rho_{10,00}\\
\rho_{01,11}&\rho_{01,10}&\rho_{01,01}&\rho_{01,00}\\
\rho_{00,11}&\rho_{00,10}&\rho_{00,01}&\rho_{00,00}
\end{pmatrix},
\end{align}
which is a positive semi-definite matrix satisfying ${\rm Tr}\rho = 1$.

\begin{figure*}[ht!]
\includegraphics[width=0.8\linewidth]{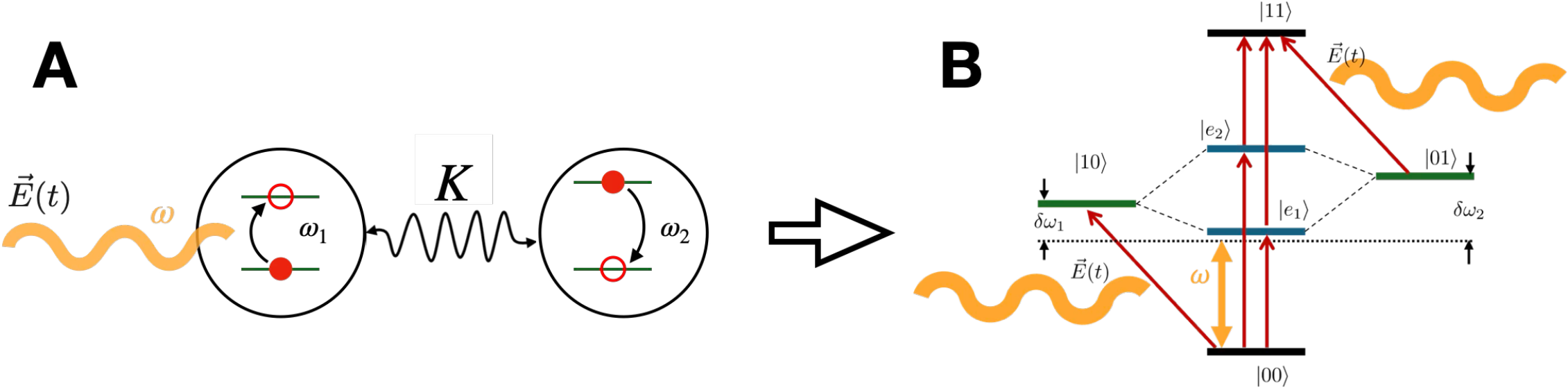}
\caption{({\bf A}) Two-qubit system coupled with a strength $K$ in a bosonic bath, one of which (qubit 1) is  irradiated by an external field with the angular frequency $\omega$. ({\bf B}) Energy diagram of two-qubit system under the eigen-basis. The singly excited states, $\ket{10}$ and $\ket{01}$ in ({\bf A}), are coupled to form two exciton states denoted by $\left|e_1\right\rangle$ and $\left|e_2\right\rangle$. 
}
\label{fig:system} 
\end{figure*}

After applying the rotating-wave approximation (RWA) {\color{black}under the condition of $\omega\gg \Omega$ along with $\omega\gg\delta\omega$, $K$, $\gamma_1$} and representing each density matrix element in a rotating frame, 
we obtain the Liouville equation for the vectorized form of the reduced density matrix in the Fock-Liouville space,   
\begin{align}
    \partial_\tau\tilde{\rho}(\tau)=\mathcal{L}\tilde{\rho}(\tau), 
    \label{eqn:Liouville_eq}
\end{align}
\begin{widetext}
\noindent where $\tilde{\rho}=(\rho_{11,11},\rho_{11,10},\rho_{11,01},\rho_{11,00},
    \rho_{10,11},\ldots,
    \rho_{01,11},\ldots,
    \rho_{00,11},\ldots
    )^T$. Eq.~\eqref{eqn:Liouville_eq} is given explicitly in SI. 
\end{widetext}

\subsection{Energy levels of TQS and resonance condition}
Due to the finite coupling strength $K$, the singly excited states (namely, $\left|01\right\rangle$ or $\left|10\right\rangle$) are no longer the eigenstates of the system Hamiltonian. 
They, instead, comprise two eigenstates, denoted by $\ket{e_{1,2}}$ (Fig.~\ref{fig:system}), and the system is described in terms of four non-degenerate eigenstates with the energy levels satisfying $E_{\ket{00}}<E_{\ket{e_1}}<E_{\ket{e_2}}<E_{\ket{11}}$ with 
\begin{align}
E_{\ket{00}}&=-\frac{\omega_1+\omega_2}{2},\nonumber\\
E_{\ket{e_1}}&=-\sqrt{K^2+\left(\frac{\omega_1-\omega_2}{2}\right)^2},\nonumber\\
E_{\ket{e_2}}&=\sqrt{K^2+\left(\frac{\omega_1-\omega_2}{2}\right)^2},\nonumber\\
E_{\ket{11}}&=\frac{\omega_1+\omega_2}{2}. 
\end{align}
Since the field is applied only to the 1st qubit (Fig.~\ref{fig:system}A),  
the permissive transitions are: 
$\ket{00}\rightarrow \ket{e_1}$, 
$\ket{00}\rightarrow \ket{e_2}$, 
$\ket{e_1}\rightarrow\ket{11}$, and 
$\ket{e_2}\rightarrow\ket{11}$ (Fig.~\ref{fig:system}B). 
Consequently, from the relations $\hbar\omega=\delta E_\pm$ with $\delta E_-=E_{\ket{e_1}}-E_{\ket{00}}=E_{\ket{11}}-E_{\ket{e_2}}$ and $\delta E_+=E_{\ket{e_2}}-E_{\ket{00}}=E_{\ket{11}}-E_{\ket{e_1}}$, the resonant frequency $\omega$ that matches with the two energy gaps between the transition-allowed eigenstates is given by 
\begin{align}
\label{eq:resonance1}
\omega_{\pm}=\frac{\omega_1+\omega_2}{2} \pm \sqrt{K^2+\left(\frac{\omega_1-\omega_2}{2}\right)^2}.  
\end{align}
The two resonance frequencies, $\omega_+$ and $\omega_-$, obtained in Eq.~\eqref{eq:resonance1} 
correspond to the two energy gaps indicated by the red arrows in Fig.~\ref{fig:system}B.  
With the definition of the detuning frequency $\delta \omega_i=\omega_i-\omega$, the two separate resonance conditions $\omega=\omega_+$ and 
$\omega=\omega_-$ can be combined through Eq.~\eqref{eq:resonance1} into 
a more succinct expression: 
\begin{align}
\delta\omega_1\delta\omega_2=K^2.
\label{eqn:resonance}
\end{align}

\subsection{Uncertainty product for TQS}
In the TQS, the uncertainty product for the steady-state TUR defined between the dissipation (total entropy production) $\Delta S_{tot}=\Sigma^{\rm ss}\tau$ for the time duration $\tau$ and the squared relative error in the output observable $n(\tau)$, specifically referring to the number of net transitions induced by the external field for the time interval $\tau$, is
\begin{align}
    \mathcal{Q}&=\lim_{\tau\rightarrow\infty}\frac{{\Sigma^{\rm{ss}}}\tau}{k_B}\frac{{\rm Var}(n (\tau))}{\left\langle n (\tau)\right\rangle^2}. 
    \label{eqn:Q}
    \end{align}
    Using the definition of photon current through the $i$-th qubit $\langle j_i\rangle=\langle n_i(\tau)\rangle/\tau$, 
    one can recast Eq.~\eqref{eqn:Q} into the product between the affinity ({\color{black}entropy production per photon emission,} $\mathcal{A}=(\Delta S_{tot}/k_B)/\langle n(\tau)\rangle$)
    and the Fano factor ($\mathcal{F}$) of photon current: 
\begin{align}
\mathcal{Q}&=\underbrace{\left(\frac{{\Sigma^{\rm{ss}}}}{k_B \langle j\rangle}\right)}_{= \mathcal{A}}\underbrace{\left(\frac{{\rm Var}(j)}{\langle j\rangle}\right)}_{= \mathcal{F}},
\label{eqn:uncertainty_product}
\end{align}
where $j=\sum_{i=1,2}j_i$. 
Here, we use the fact that the total entropy production from the system is contributed by the net number of photon emissions that have occurred from both qubits, namely, 
$\Sigma^{\rm ss}\tau=\mathcal{A}_1\langle n_1(\tau)\rangle+\mathcal{A}_2\langle n_2(\tau)\rangle$, with $\mathcal{A}_i=\hbar\omega_i/k_BT$, which thus yields 
$\mathcal{A}=\frac{\sum_{i=1,2}\mathcal{A}_i\langle j_i\rangle}{ \sum_{i=1,2}\langle j_i\rangle}$. 
The mean current $(\left\langle j \right\rangle)$ and the current fluctuations $({\rm Var}(j))$ at steady states are calculated using the method of cumulant generating function (see Methods).

\subsection{Dynamic properties of the field-driven dissipative TLS}
If the coupling strength is set to zero ($K=0$), the 1st qubit is fully decoupled from the 2nd qubit. Thus, the expressions for the mean current and fluctuations are reduced to those of the TLS subject to an external field~\cite{singh2021}. 
Since the expressions of current and current fluctuations for the field-driven dissipative TLS, corresponding to TQS with $K=0$, are of great use for our subsequent interpretation of TQS, we write them down explicitly. 
\begin{align}
    \langle j\rangle=\frac{\Omega^2}{2[\Omega^2+\Omega_o^2+\delta\omega^2/2]}
    \label{eqn:j}
\end{align}
with $\Omega_o=\frac{1}{2\sqrt{2}}\coth{(\mathcal{A}/2)}$, 
and 
\begin{align}
    {\rm Var}(j)=\langle j\rangle\coth{\left(\frac{\mathcal{A}}{2}\right)}\left(1+2\rho_R^2-6\rho_I^2\right). 
        \label{eqn:Var_j}
    \end{align}
Thus, the uncertainty product for the field-driven dissipative TLS is written as 
    \begin{align}
\mathcal{Q}^{\rm TLS}&=\mathcal{A}\frac{{\rm Var}(j)}{\langle j\rangle}\nonumber\\
&=\mathcal{A}\coth{\left(\frac{\mathcal{A}}{2}\right)}\left(1+2\rho_R^2-6\rho_I^2\right). 
\label{eqn:Q_TLS}
\end{align}
Here, $\rho_R$ and $\rho_I$ are the real and imaginary part of the coherence between the excited and ground states ($\rho_{10}^{\rm ss}=\rho_R+i\rho_I$) at steady states, respectively, defined in the rotating frame.
The explicit expressions are given as follows~\cite{singh2021}:
\begin{align}
\rho_R=\frac{-\Omega\delta\omega}{2\coth{\left(\frac{\mathcal{A}}{2}\right)}\left[\Omega^2+\Omega_o^2+\delta\omega^2/2\right]}
\label{eqn:rhoR}
\end{align}
and 
\begin{align}
\rho_I
=\frac{\Omega}{4\left[\Omega^2+\Omega_o^2+\delta\omega^2/2\right]}. 
\label{eqn:rhoI}
\end{align}

A couple of key remarks are in place. 

(i) The \emph{current-coherence relation} at nonequilibrium steady states~\cite{wu2012efficient,yang2020steady} is obtained using Eq.~\eqref{eqn:j} and Eq.~\eqref{eqn:rhoI}.  
\begin{align}
\langle j\rangle=2\Omega\rho_I. 
\label{eqn:j_rhoI}
\end{align}
The relation states that the imaginary part of the coherence is responsible for the steady state current generated by quantum transition. 
As physically anticipated, 
the Rabi frequency ($\Omega$), which arises from the external driving, plays a key role not only in generating the photon current by facilitating the quantum transitions of the TLS, but also in sustaining the coherence. 
Without external driving ($\Omega=0$), the mean current (Eq.~\eqref{eqn:j}), current fluctuations (Eq.~\eqref{eqn:Var_j}), and coherences (Eqs.~\eqref{eqn:rhoR} and ~\eqref{eqn:rhoI}) vanish altogether, and one obtains $\mathcal{Q}^{\rm TLS}=\mathcal{A}\coth{\left(\frac{\mathcal{A}}{2}\right)}\xrightarrow{\mathcal{A}\rightarrow0}2$.

(ii) At perfect resonance ($\delta\omega=0$), $\rho_R^2=0$, and $\rho_I^2$, unimodal with respect to $\Omega$, is maximized to 
$(\rho_I^2)_{\rm max}=\frac{1}{8\coth^2{\left(\frac{\mathcal{A}}{2}\right)}}$
at $\Omega=\Omega_o$. 
Thus, 
\begin{align}
\mathcal{Q}^{\rm TLS}\geq \mathcal{A}\coth{\frac{\mathcal{A}}{2}}\left(1-\frac{3}{4\coth^2{\left(\frac{\mathcal{A}}{2}\right)}}\right)\geq  \mathcal{Q}_{\rm min}^{\rm TLS}. 
\label{eqn:Qmin}
    \end{align}
The minimal TUR bound for the field-driven TLS is $\mathcal{Q}_{\rm min}^{\rm TLS}\approx 1.24_6$, 
which is obtained when $\mathcal{A}\approx 3.61_0$. 

\begin{figure*}[ht!]
\includegraphics[width=0.8\linewidth]{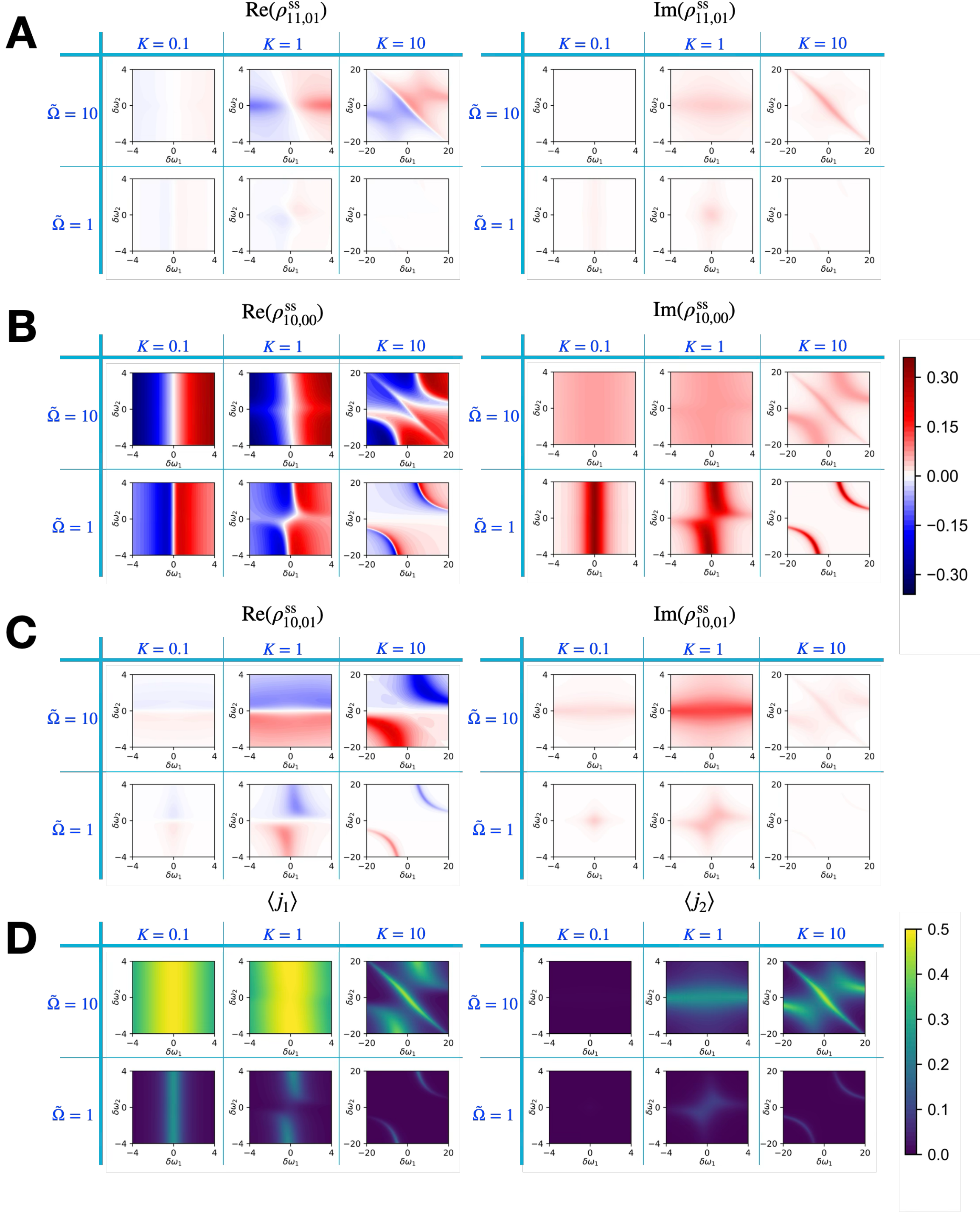}
\caption{Real and imaginary parts of coherence of the 1st qubit when the 2nd qubit is in its (A) excited and (B) ground states. (C) The correlated coherence. 
(D) Mean photon current generated from the 1st and 2nd qubits. 
The coherences and current for $\tilde{\Omega}=0.1$ are not explicitly shown since they are nearly zeros. The results are obtained for $\mathcal{A}_1=\mathcal{A}_2=3.6$. 
}
\label{fig:current_coherence} 
\end{figure*}

\section{Results}
\subsection{Populations and coherences of TQS and current-coherence relations}
Without external driving ($\Omega=0$), 
the density matrix at steady states has a form of
\begin{align}
\rho^{\textrm{ss}} = 
\begin{pmatrix}
\rho_{11,11}^{\textrm{ss}}& 0 & 0 & 0\\
0 &\rho_{10,10}^{\textrm{ss}}&\rho_{10,01}^{\textrm{ss}}& 0 \\
0 &\rho_{01,10}^{\textrm{ss}}&\rho_{01,01}^{\textrm{ss}}& 0 \\
0 & 0 &  0& \rho_{00,00}^{\textrm{ss}} 
\end{pmatrix}, 
\end{align}
which is expected from the physical constraint and symmetry imposed on our problem. 
The population at each state, i.e., the diagonal elements of the density matrix, is obtained as
\begin{align}
\label{eqn:populations}
    \rho^{\textrm{ss}}_{11,11}&= 
    [K^2(\nbar_1+\nbar_2)^2+\nbar_1\nbar_2q]/\mathcal{D}\nonumber\\
    \rho^{\textrm{ss}}_{10,10}&=
    [K^2(\nbar_1+\nbar_2)(\nbar_1+\nbar_2+2)+\nbar_1(1+\nbar_2)q]/\mathcal{D}\nonumber\\
    \rho^{\textrm{ss}}_{01,01}&=
    [K^2(\nbar_1+\nbar_2)(\nbar_1+\nbar_2+2)+(1+\nbar_1)\nbar_2q]/\mathcal{D}\nonumber\\
    \rho^{\textrm{ss}}_{00,00}&=[K^2(\nbar_1+\nbar_2+2)^2+(1+\nbar_1)(1+\nbar_2)q]/\mathcal{D},
    \end{align}
    where 
    $q\equiv \left[(\delta\omega_1-\delta\omega_2)^2+(\nbar_1+\nbar_2+1)^2\right]$ and 
    $\mathcal{D}\equiv 4K^2(\nbar_1+\nbar_2+1)^2+(1+2\nbar_1)(1+2\nbar_2)q$. 
Together with 
$\nbar_{1,2}=(e^{\mathcal{A}_{1,2}}-1)^{-1}\ll 1$ for $\mathcal{A}_{1,2}\approx 3.61$, 
the excited-to-ground state population ratio at weak coupling ($K\ll1$)  
satisfies the Boltzmann distribution for the 1st qubit 
\begin{align}
\frac{\rho^{\rm ss}_{11,11}}{\rho^{\rm ss}_{01,01}}, \frac{\rho^{\rm ss}_{10,10}}{\rho^{\rm ss}_{00,00}}\simeq \frac{\nbar_1}{\nbar_1+1}=e^{-\beta\hbar\omega_1}
\label{eqn:smallK1}
\end{align}
and for the 2nd qubit
\begin{align}
\frac{\rho^{\rm ss}_{01,01}}{\rho^{\rm ss}_{00,00}}, \frac{\rho^{\rm ss}_{11,11}}{\rho^{\rm ss}_{10,10}}\simeq \frac{\nbar_2}{\nbar_2+1}=e^{-\beta\hbar\omega_2}. 
\label{eqn:smallK2}
\end{align}
For $K\gg 1$, on the other hand, 
all the ratios are approximated to the same expression,
\begin{align}
\frac{\rho^{\rm ss}_{11,11}}{\rho^{\rm ss}_{01,01}}, \frac{\rho^{\rm ss}_{10,10}}{\rho^{\rm ss}_{00,00}},\frac{\rho^{\rm ss}_{01,01}}{\rho^{\rm ss}_{00,00}}, \frac{\rho^{\rm ss}_{11,11}}{\rho^{\rm ss}_{10,10}}\simeq\frac{(\nbar_1+\nbar_2)}{(\nbar_1+\nbar_2+2)}. 
\label{eqn:largeK}
\end{align} 

Next, quantum coherences, quantified via off-diagonal elements of the density matrix, were previously linked to the TUR violation of the field-driven dissipative TLS~\cite{singh2021}.
Due to the hermicity of the
density matrix ($\rho_{ij,kl}=\rho_{kl,ij}^\ast$), 
the coherences of the TQS studied here are fully characterized
by six distinct off-diagonal elements, $\rho^{\rm ss}_{11,10}$, $\rho^{\rm ss}_{11,01}$, $\rho^{\rm ss}_{11,00}$, $\rho^{\rm ss}_{10,01}$, $\rho^{\rm ss}_{10,00}$, and $\rho^{\rm ss}_{01,00}$. 
Specifically, 
$\rho^\textrm{ss}_{10,00}$
and  $\rho^\textrm{ss}_{11,01}$
represent 
the coherence of the 1st qubit while the 2nd qubit is in its ground and excited states, respectively; 
$\rho^\textrm{ss}_{01,00}$ and  $\rho^\textrm{ss}_{11,10}$ are for the coherence of the 2nd  qubit; 
$\rho^\textrm{ss}_{10,01}$ and $\rho^\textrm{ss}_{11,00}$ are 
interpreted as  the \emph{correlated coherences} between the two qubits. 

It is expected from the two-qubit coupling hamiltonian (Eq.~\eqref{eqn:coupling_H}) that a strong coupling ($K\gg 1$) is required to efficiently transmit excitation
of the 1st atom to the 2nd atom and to generate correlation between the two qubits.
This is generally true, but only if the intensity of the field ($\Omega$) is large enough to 
induce excitation and coherence
in the 1st qubit. 
Otherwise, neither is formed the coherence within an atom ($\rho_R\approx 0$, $\rho_I\approx 0$ from Eqs.~\eqref{eqn:rhoR} and ~\eqref{eqn:rhoI}) nor the correlation between the two atoms. 

Fig.~\ref{fig:current_coherence} demonstrates the coherences and currents of the TQS under varying amount of detunings ($\delta\omega_1$, $\delta\omega_2$) for select values of Rabi frequency ($\Omega/\Omega_o=0.1$, 1, 10) and coupling strength ($K=0.1, 1, 10$). 
For the sake of our discussion, 
we normalize the Rabi frequency by $\Omega_o=0.37$ that minimizes $\mathcal{Q}^{\rm TLS}$ (Eq.~\eqref{eqn:Qmin})~\cite{singh2021}, defining $\tilde{\Omega}\equiv \Omega/\Omega_o$. 
As commented in Eq.~\eqref{eqn:j_rhoI}, 
the photon current generated in an isolated qubit interacting with an external field is directly proportional to the imaginary part of coherence~\cite{wu2012efficient,yang2020steady,roden2016probability}. 
This situation arises when the 1st qubit is subject to the external field, whereas the 2nd qubit is effectively decoupled from the 1st qubit due to the weak coupling ($K\ll 1$).  
${\rm Im}(\rho_{10,00}^{\rm ss})$ for three $\tilde{\Omega}$ values at $K=0.1$ (the 
first column of Fig.~\ref{fig:current_coherence}B), reflecting 
the functional form of Eq.~\eqref{eqn:rhoI} that maximizes at $\tilde{\Omega}=1$ (or $\Omega=\Omega_o$),   
directly translate to the mean current generated in the 1st qubit 
($\langle j_1\rangle$, the first column of   Fig.~\ref{fig:current_coherence}B). 
It is also noteworthy that under the same condition ($K=0.1$), 
the coherence (imaginary part) of the 2nd qubit is effectively zero (${\rm Im}[\rho_{01,00}^{\rm ss}]\approx 0$. See Fig.~S1B). 

The full expressions for the current, current fluctuations, and density matrix elements of TQS at steady states are too extensive to be shown here explicitly. 
However, either by noting that the photon current from each qubit is 
physically generated via \emph{two distinct channels} of transitions 
or by using 
the general expression of the current (see Eq.~\eqref{eqn:current_general} {\color{black}derived in Methods}), 
$\langle j_1\rangle$ and $\langle j_2\rangle$ are written in terms of the difference between the steady state populations, 
\begin{widetext}
\begin{align}
\langle j_1\rangle&=[(\nbar_1+1)\rho_{11,11}^{\rm ss}-\nbar_1\rho_{01,01}^{\rm ss}]+[(\nbar_1+1)\rho_{10,10}^{\rm ss}-\nbar_1\rho_{00,00}^{\rm ss}], 
\label{eqn:j1_explicit}
\end{align}
\begin{align}
\langle j_2\rangle&=[\gamma(\nbar_2+1)\rho_{11,11}^{\rm ss}-\gamma\nbar_2\rho_{10,10}^{\rm ss}]+[\gamma(\nbar_2+1)\rho_{01,01}^{\rm ss}-\gamma\nbar_2\rho_{00,00}^{\rm ss}]. 
\label{eqn:j2_explicit}
\end{align}
Note that Eq.~\eqref{eqn:j1_explicit}, for instance, expresses the two channels of photon emission from the 1st qubit while the 2nd qubit is in its excited and ground state. 
In fact, one can also relate the currents with the imaginary parts of coherences, establishing 
the current-coherence relations for TQS as follows (see Eqs.~(S2) --~(S5) in SI for the detailed derivation):  
\begin{align}
\langle j\rangle=2\Omega({\rm Im}[\rho_{11,01}^{\rm ss}]+{\rm Im}[\rho_{10,00}^{\rm ss}]), 
\label{eqn:jtot}
\end{align}
\begin{align}
    \left\langle j_1\right\rangle 
    &=
    2 \Omega \left(\rm{Im}\left[\rho^{\rm{ss}}_{11,01}\right]+\rm{Im}\left[\rho^{\rm{ss}}_{10,00}\right]\right)-2 K \rm{Im}\left[\rho^{\rm{ss}}_{10,01}\right], 
    \label{eqn:j1}
    \end{align}
    \begin{align}
    \left\langle j_2\right\rangle 
    &=
    2 K \rm{Im}\left[\rho^{\rm{ss}}_{10,01}\right]. 
    \label{eqn:j2}
\end{align}
\end{widetext}
The physical setup of TQS -- only the 1st qubit is directly irradiated by the external field, whereas the 2nd qubit is indirectly influenced by the external field through the coupling with the 1st qubit -- is reflected in the expressions of  Eqs.~\eqref{eqn:jtot},~\eqref{eqn:j1}, and~\eqref{eqn:j2}. 
The total current from the TQS (Eq.~\eqref{eqn:jtot}) is similar to the current-coherence relation for the TLS (Eq.~\eqref{eqn:j_rhoI}) in that it depends only on the imaginary parts of the coherences of the 1st qubit, ${\rm Im}[\rho_{11,01}^{\rm ss}]$ and ${\rm Im}[\rho_{10,00}^{\rm ss}]$, with dominant contributions from ${\rm Im}[\rho_{10,00}^{\rm ss}]$. 
The current from the 2nd qubit, $\langle j_2\rangle$, is determined solely by the correlated coherence, $\rm{Im}\left[\rho^{\rm{ss}}_{10,01}\right]$, 
with the coupling strength $K$ as the proportionality constant. 
Notably, the \emph{even} centrosymmetries of $\langle j_{1,2}\rangle$ with respect to $\delta\omega_{1,2}$ demonstrated in Fig.~\ref{fig:current_coherence}D, i.e., $\langle j_{1,2}\rangle(\delta\omega_1,\delta\omega_2)=\langle j_{1,2}\rangle(-\delta\omega_1,-\delta\omega_2)$ are consistent only with those of ${\rm Im}[\rho_{11,01}^{\rm ss}]$ (Fig.~\ref{fig:current_coherence}A), ${\rm Im}[\rho_{10,00}^{\rm ss}]$ (Fig.~\ref{fig:current_coherence}B), and ${\rm Im}[\rho_{10,01}^{\rm ss}]$ (Fig.~\ref{fig:current_coherence}C).

\begin{figure*}[ht!]
\includegraphics[width=0.8\linewidth]{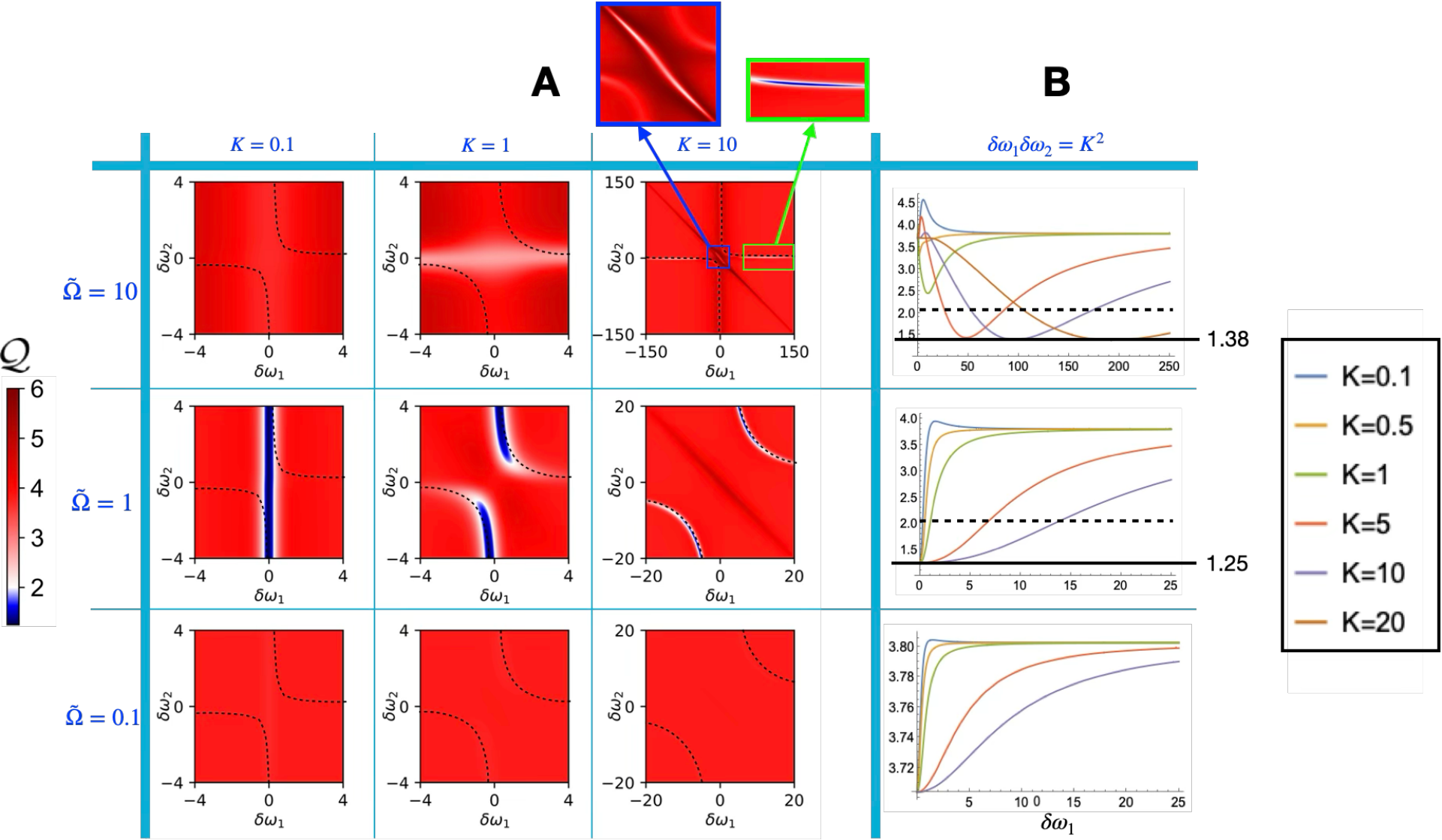}
\caption{({\bf A}) 2D map of uncertainty product over varying extent of detunings, $\mathcal{Q}(\delta\omega_1,\delta\omega_2)$, with $\mathcal{A}_{1,2}=3.6$ for $\tilde\Omega=0.1$, 1, 10 and $K=0.1$, 1, 10. 
({\bf B}) $\mathcal{Q}(\delta\omega_1)$ along the resonance condition ($\delta\omega_1\delta\omega_2=K^2$) marked with the curved dashed lines in ({\bf A}). 
The dashed lines mark $\mathcal{Q}=2$. The solid lines annotated with 1.38 and 1.25 denote the minimum value of $\mathcal{Q}$.
}
\label{fig:Q_detuning} 
\end{figure*}

\subsection{The uncertainty product of two-qubit system}
Together with ${\rm Var}(j)(\delta\omega_1,\delta\omega_2)$ (Fig.~S2) and $\langle j\rangle(\delta\omega_1,\delta\omega_2)$ (Fig.~\ref{fig:current_coherence}D) calculated at fixed $\mathcal{A}=3.61$,  
we use Eq.~\eqref{eqn:uncertainty_product} to obtain $\mathcal{Q}(\delta\omega_1,\delta\omega_2)$ for TQS as a function of $\tilde{\Omega}$ and $K$. 
Here are our findings from Fig.~\ref{fig:Q_detuning}. 

(i) For weak driving  $\tilde{\Omega}=0.1$, the value of $\mathcal{Q}$ is nearly constant over varying amount of detunings, regardless of $K$.
This is understood since the system is effectively unperturbed, remaining at equilibrium.

(ii) In Fig.~\ref{fig:Q_detuning}A, the condition of $(\delta\omega_1,\delta\omega_2)$ leading to the TUR violation ($\mathcal{Q}<2$) is the most evident  
when $\tilde{\Omega}\approx 1$. 
For weak coupling between the qubits ($K=0.1$), the $\mathcal{Q}<2$ is found for $\delta\omega_1\approx 0$, which is tantamount to the resonance condition for an isolated qubit (TLS). 
For $K=1$, 
the range of $(\delta\omega_1,\delta\omega_2)$ involving $\mathcal{Q}<2$ is aligned along the resonance condition of TQS, $\delta\omega_1\delta\omega_2=K^2$ (Eq.~\eqref{eqn:resonance}), particularly at $|\delta\omega_1|<1$ and $|\delta\omega_2|>1$. 
For $K=10$, $\mathcal{Q}(\delta\omega_1,\delta\omega_2)<2$ even better aligns with the TQS resonance condition.

(iii) For stronger driving ($\tilde{\Omega}=10$), the region of TUR violation ($\mathcal{Q}<2$) is found along the resonance condition with $|\delta\omega_1|\gg 1$ and $|\delta\omega_2|\ll1$ when $K=10$. 
Although it is not involved with the TUR violation, a region with small $\mathcal{Q}(\gtrsim 2)$ is identified along the condition of $\delta\omega_1=-\delta\omega_2$ (see Fig.~\ref{fig:Q_detuning}A inset enclosed by the blue square).

(iv) Fig.~\ref{fig:Q_detuning}B, plotting the value of $\mathcal{Q}$ as a function of $\delta\omega_1$ by replacing $\delta\omega_2$ with $K^2/\delta\omega_1$, makes more explicit 
the range of $\delta\omega_1$ leading to the TUR violation 
along the resonance condition ($\delta\omega_1\delta\omega_2=K^2$). 
For $\tilde{\Omega}=1$ (the panel in the middle of Fig.~\ref{fig:Q_detuning}B), the $\delta\omega_1$ giving rise to the loose TUR bound $\mathcal{Q}_{\rm min}\approx 1.25<2$ is found around $0<\delta\omega_1<1$. 
The corresponding region broadens with increasing $K$. 
For $\tilde{\Omega}=10$, on the other hand, 
$\delta\omega_1$'s giving rise to the loose TUR bound are identified 
around $\delta\omega_1\approx 50$, 100, 190 for $K=5$, 10, and 20, respectively. 
It is noteworthy that $\mathcal{Q}_{\rm min}\approx 1.38$ for $\tilde{\Omega}=10$.



\begin{figure}[t]
\includegraphics[width=0.9\linewidth]{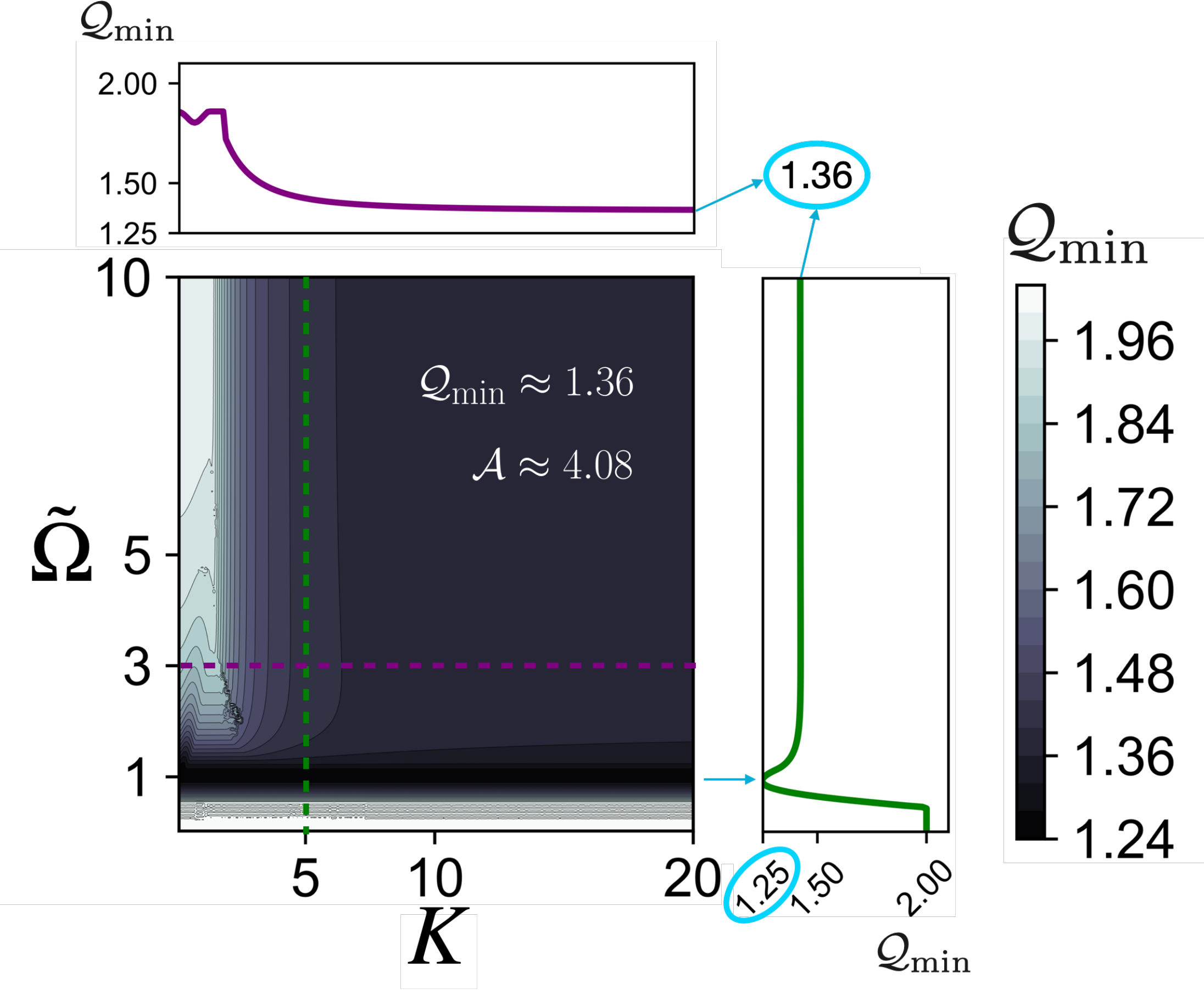}
\caption{
The 2D map of the minimal bound of TUR, $\mathcal{Q}_{\rm min}(K, \tilde{\Omega})$ obtained along the resonance condition ($\delta\omega_1\delta\omega_2=K^2$) while varying $\delta\omega_1$ and $\mathcal{A}$. 
The cross-sectional profiles of $\mathcal{Q}_{\rm min}$, $\mathcal{Q}_{\rm min}(K,\tilde{\Omega}=3)$ and $\mathcal{Q}_{\rm min}(K=5,\tilde{\Omega})$ are plotted on the right and at the top, respectively. Highlighted by the cyan circles are the two distinct values of TUR bound for TQS, $\mathcal{Q}_{\rm min}\simeq1.25$ and $1.36$ obtained under the conditions of $\tilde{\Omega}\approx 1$ ($\mathcal{A}\approx 3.61$) and $(\tilde{\Omega}\gtrsim 3,K\gtrsim 10)$ ($\mathcal{A}\approx 4.08$).
}
\label{fig:okmin} 
\end{figure}

\section{Discussions}
\subsection{Minimal value of the uncertainty product, $\mathcal{Q}_{\rm min}$.} 
From the 2D map of $\mathcal{Q}(\delta\omega_1,\delta\omega_2)$ (Fig.~\ref{fig:Q_detuning}A) and $\mathcal{Q}(\delta\omega_1)$ along the resonance condition (Fig.~\ref{fig:Q_detuning}B) for select values of $\tilde{\Omega}$ and $K$, 
we find that $\mathcal{Q}$ is minimized along the resonance condition, which reads $\delta\omega_1=0$ for small $K(\ll 1)$ and $\delta\omega_1\delta\omega_2=K^2$ for large $K(\gtrsim 1)$. 
At least two distinct values of $\mathcal{Q}_{\rm min}=1.38$  and 1.25, which are smaller than the classical counterpart ($\mathcal{Q}_{\rm min}=2$), are highlighted in Fig.~\ref{fig:Q_detuning}B. 
To identify the value of $\mathcal{Q}_{\rm min}$ 
along the resonance condition more systematically, 
we calculate the diagram of $\mathcal{Q}_{\rm min}$ as a function of $\tilde{\Omega}$ and $K$ (Fig.~\ref{fig:okmin}) by simultaneously scanning $\delta\omega_1$ and $\mathcal{A}$.

In case of weak-to-moderate driving ($\tilde{\Omega}\lesssim 1$), $\mathcal{Q}_{\rm min}$ is nearly insensitive to the coupling strength ($K$). 
The uncertainty product minimizes to $\mathcal{Q}_{\rm min}\approx1.25$ at $\tilde{\Omega}\approx 1$.
In fact,  $\mathcal{Q}_{\rm min}\approx1.25$ is obtained 
when $\delta\omega_1\approx 0$ (Fig.~\ref{fig:Q_detuning}B, the middle panel).
Since $\delta\omega_2=K^2/ \delta\omega_1$, the condition of $\delta\omega_1\approx 0$ signifies $\delta\omega_2\rightarrow\infty$, effectively decoupling the 2nd qubit from the system. 
In this limit, the TQS behaves effectively like a TLS, and thus it is not surprising that the TUR bound converges to that of a TLS. \cite{singh2021}.

For strong driving $\tilde{\Omega}\gg 1$ and weak coupling limit ($K\ll 1$), 
the system reduces to the TLS, attaining $\mathcal{Q}_{\rm min}\approx 2$ for $\mathcal{A}\approx0$~\cite{singh2021}. 
On the other hand, $\mathcal{Q}_{\rm min}$ forms a broad plateau of $\mathcal{Q}_{\rm min}\approx 1.36$ at $K\gtrsim 5$ and $\tilde{\Omega}\gtrsim 3$. 
We, in fact, find that although the expression of $\mathcal{Q}$ is too extensive to show, 
$\mathcal{Q}$ is minimized to $\mathcal{Q}_{\rm min}=1.35_8$ when $\mathcal{A}_1\approx 4.08$ 
under a condition of $\delta\omega_1=K\tilde{\Omega}$ and $\delta\omega_2=K/\tilde{\Omega}$ (see Fig.~S3).  

Taken together, 
the minimal bound of the uncertainty product for the TQS, $\mathcal{Q}_{\rm min}\approx 1.25$, is acquired as long as the field is moderate ($\tilde{\Omega}\approx 1$).
Since $\mathcal{Q}_{\rm min}\approx 1.25$ is identical to that of a single TLS, 
the TQS effectively responds to the external field like a field-driven single TLS, even in the presence of the 2nd qubit coupled to the 1st qubit at finite $K$. 
For $\tilde{\Omega}\gg 1$ and $K\gg 1$, 
we find that the TUR of TQS is characterized with another value, $\mathcal{Q}_{\rm min}\approx 1.36$. 

\begin{figure}[t]
\includegraphics[width=1.0\linewidth]{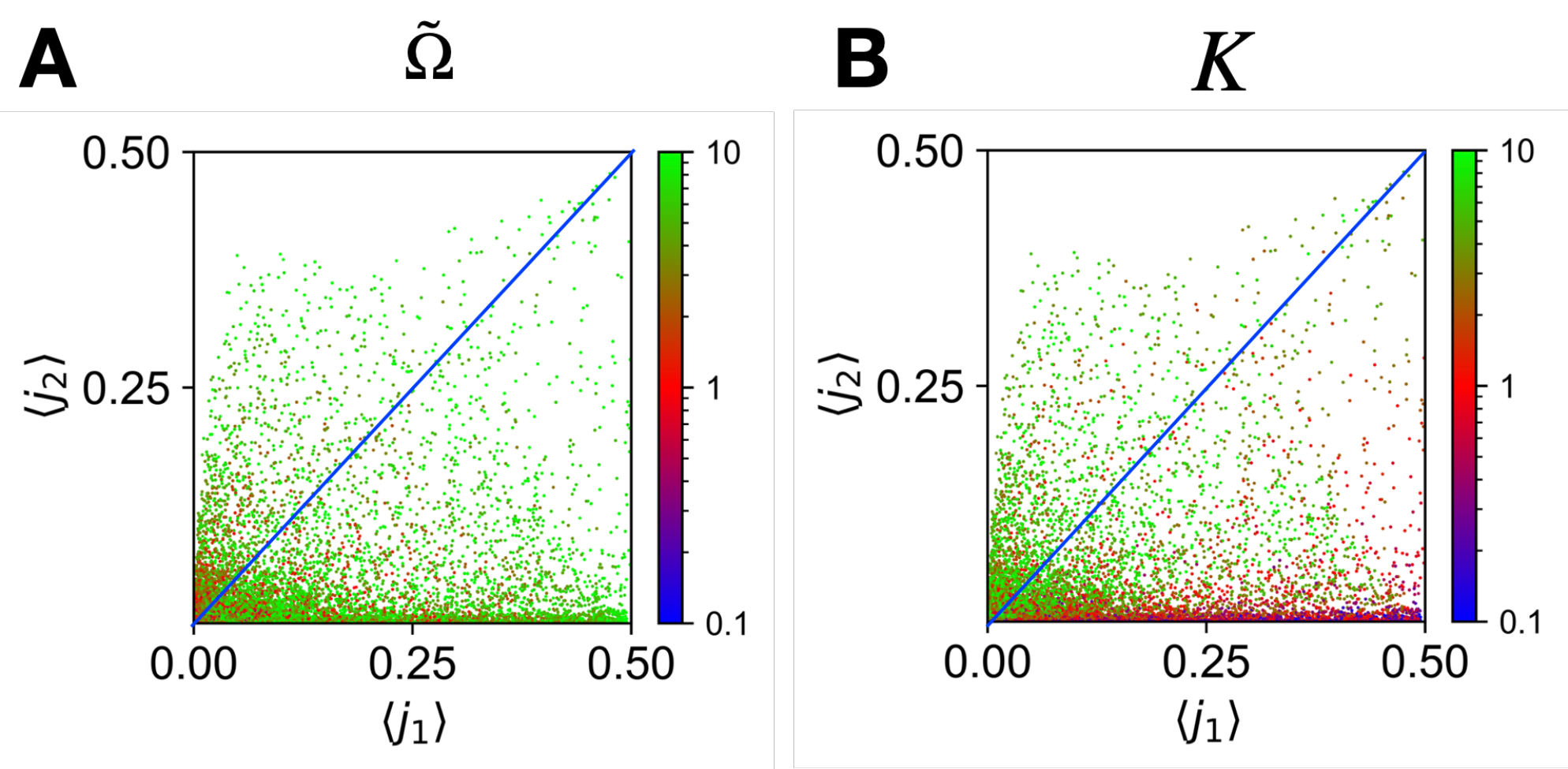}
\caption{Asymmetry of the currents from the two qubits. 
({\bf A}) $\tilde{\Omega}(\langle j_1\rangle,\langle j_2\rangle)$ obtained from randomly sampled set of parameters ($K$, $\delta\omega_1$, $\delta\omega_2$) with $\mathcal{A}_{1,2}=3.6$. 
({\bf B}) $K(\langle j_1\rangle,\langle j_2\rangle)$ obtained from ($\tilde{\Omega}$, $\delta\omega_1$, $\delta\omega_2$) with $\mathcal{A}_{1,2}=3.6$.  
The blue lines along the diagonal are a guide for the eye. 
}
\label{fig:curr_2d} 
\end{figure}

\subsection{The asymmetry of two qubits.}
The coherences of the 2nd qubit inherently differ from those of the 1st qubit. 
Specifically, for the 1st qubit, the real and imaginary parts of the coherences, respectively, exhibit odd and even centrosymmetry under the inversion from $(\delta\omega_1,\delta\omega_2)$ to $(-\delta\omega_1,-\delta\omega_2)$  (Fig.~\ref{fig:current_coherence}A and ~\ref{fig:current_coherence}B), i.e., 
\begin{align}
{\rm Re}[\rho^{\rm ss}_{\rm 1}](\delta\omega_1,\delta\omega_2)&=-{\rm Re}[\rho^{\rm ss}_{\rm 1}](-\delta\omega_1,-\delta\omega_2)\nonumber\\
{\rm Im}[\rho^{\rm ss}_{\rm 1}](\delta\omega_1,\delta\omega_2)&={\rm Im}[\rho^{\rm ss}_{\rm 1}](-\delta\omega_1,-\delta\omega_2),
\end{align}
where $\rho^{\rm ss}_{\rm 1}$ denotes either $\rho^{\rm ss}_{11,01}$ or $\rho^{\rm ss}_{10,00}$. 
For the 2nd qubit, on the other hand, 
\begin{align}
{\rm Re}[\rho^{\rm ss}_{\rm 2}](\delta\omega_1,\delta\omega_2)&={\rm Re}[\rho^{\rm ss}_{\rm 2}](-\delta\omega_1,-\delta\omega_2)\nonumber\\
{\rm Im}[\rho^{\rm ss}_{\rm 2}](\delta\omega_1,\delta\omega_2)&=-{\rm Im}[\rho^{\rm ss}_{\rm 2}](-\delta\omega_1,-\delta\omega_2),
\end{align}
where $\rho^{\rm ss}_{\rm 2}$ denotes either $\rho^{\rm ss}_{11,10}$ or $\rho^{\rm ss}_{01,00}$ 
(Fig.~S1A and ~S1B). 
Lastly, the centrosymmetry of the correlated coherences is even for both 
$\rho^{\rm ss}_{11,00}$ and $\rho^{\rm ss}_{10,01}$, such that   
\begin{align}
{\rm Re}[\rho^{\rm ss}_{\rm corr}](\delta\omega_1,\delta\omega_2)&={\rm Re}[\rho^{\rm ss}_{\rm corr}](-\delta\omega_1,-\delta\omega_2)\nonumber\\
{\rm Im}[\rho^{\rm ss}_{\rm corr}](\delta\omega_1,\delta\omega_2)&={\rm Im}[\rho^{\rm ss}_{\rm corr}](-\delta\omega_1,-\delta\omega_2). 
\end{align}

In the physical setup of the TQS, where only the 1st qubit 
is subject to the external driving, 
the symmetry of the two qubits is broken, which is also reflected in the two distinct expressions of the current flowing through the two qubits, $\langle j_1\rangle$ and $\langle j_2\rangle$ derived in 
Eqs~\eqref{eqn:j1} and \eqref{eqn:j2}. 
The current generated from the 1st qubit is, by and large, greater than that from the 2nd qubit ($\langle j_1\rangle>\langle j_2\rangle$) as displayed in Fig.~\ref{fig:curr_2d}. 

Note that $\langle j_1\rangle$ is dictated by the typical current-coherence relation~(Eq.~\eqref{eqn:j_rhoI}), whereas $\langle j_2\rangle$ is associated only with the imaginary part of the correlated coherence, ${\rm Im}[\rho^{\rm ss}_{10,01}]$. 
It is noteworthy that another correlated coherence ${\rm Im}[\rho^{\rm ss}_{11,00}]$ (Fig.~S1C) makes no contribution to the current. 
This is most likely associated with the fact that 
$\langle j_2\rangle>0$ (Fig.~\ref{fig:current_coherence}D) for all $\delta\omega_1$ and $\delta\omega_2$ is more consistent with ${\rm Im}[\rho^{\rm ss}_{10,01}]$ (Fig.~\ref{fig:current_coherence}C) than 
${\rm Im}[\rho^{\rm ss}_{11,00}]$, the latter of which is not always positive (Fig.~S1C). 
Despite the inherent asymmetry, the currents generated from the 1st and 2nd qubits become, {\color{black}on average,} comparable when the two qubits are strongly coupled at large $K$ (data points in green, Fig.~\ref{fig:curr_2d}B).

\subsection{TUR of subsystem and comparison of TQS with two coupled noisy oscillators}
Throughout this study, we select the total photon current as the observable of interest. 
However, if accessible only to the dynamics of a single qubit, then 
the observable of our choice becomes the local current from a particular qubit, and one has to consider the TUR of subsystem. 

Before proceeding further, it is worth noting that the TQS studied here bears similarity, albeit classical, with two noisy oscillators, whose phase dynamics can be synchronized due to the inter-oscillator coupling with a strength $K_{os}$~\cite{lee2018PRE,sakaguchi1988PTP}:
 \begin{align}
 \frac{d\theta_1}{dt}&=\Omega_1+\frac{K_{os}}{2}\sin{(\theta_2-\theta_1)}+\xi_1(t)\nonumber\\
  \frac{d\theta_2}{dt}&=\Omega_2+\frac{K_{os}}{2}\sin{(\theta_1-\theta_2)}+\xi_2(t)
 \end{align}
where $\Omega_{1,2}$ are the angular frequencies of the two oscillators, $\xi_{1,2}$ are the noise of the bath thermalized at the temperature $T$, satisfying $\langle\xi_{1,2}(t)\rangle=0$ and $\langle \xi_i(t)\xi_j(t)\rangle=2D\delta_{ij}\delta(t-t')$ with $D=k_BT/\zeta$ and $\zeta$ the friction coefficient. 
In this problem, the uncertainty product of a subsystem is given by 
\begin{align}
\mathcal{Q}_i=\frac{\Delta S_{tot}}{k_B}\frac{\langle\delta\theta_i(\tau)^2\rangle}{\langle\theta_i(\tau)\rangle}. 
\end{align}

\begin{figure}[t]
\includegraphics[width=1\linewidth]{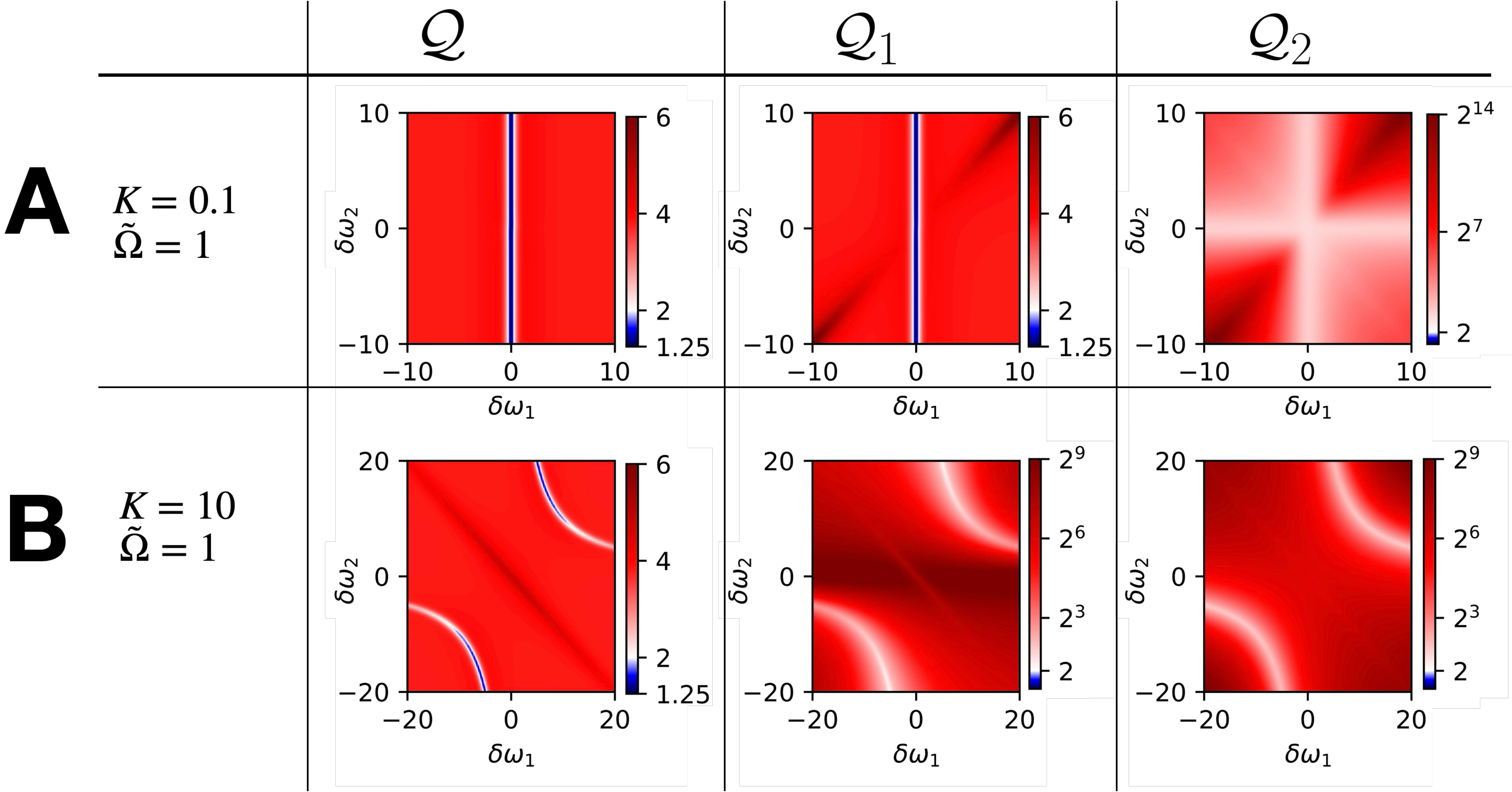}
\caption{Uncertainty products of the TQS ($\mathcal{Q}$) and each qubit ($\mathcal{Q}_1$ and $\mathcal{Q}_2$) as a function of detunings ($\delta\omega_1, \delta\omega_2$). 
({\bf A}) $K=0.1$, $\tilde{\Omega}=1$. 
({\bf B}) $K=10$, $\tilde{\Omega}=1$.}
\label{fig:partialQ} 
\end{figure}

(i) For $K_{os}=0$, the two oscillators exhibit phase dynamics fully decoupled from each other~\cite{lee2018PRE}. 
As a result, the total entropy production, the variance and the mean phase angle for $\tau\gg 1$ from each oscillator ($i=1$ and 2) are given by 
\begin{align}
\Delta S_{tot}/k_B&=\frac{\zeta(\Omega_1^2+\Omega_2^2)}{k_BT}\tau,\nonumber\\
\langle \delta\theta_i(\tau)^2\rangle&=2D\tau,\nonumber\\
\langle \theta_i(\tau)\rangle&=\Omega_i\tau. 
\end{align}
Thus, 
the uncertainty product for the $i$th oscillator is 
\begin{align}
\mathcal{Q}_i=2\left(\frac{\Omega_1^2+\Omega_2^2}{\Omega_i^2}\right)
\end{align}

(ii) For $K_{os}\gg |\Omega_1-\Omega_2|$, the phase dynamics of the two oscillators are synchronized, slowing down the fast oscillator and speeding up the slow oscillator and reaching the angular velocity of $(\Omega_1+\Omega_2)/2$~\cite{lee2018PRE,hong2020JSM}.   
Thus, we obtain the following quantities of interest:  
\begin{align}
\Delta S_{tot}/k_B&=\frac{2\zeta}{k_BT}\left(\frac{\Omega_1+\Omega_2}{2}\right)^2\tau,\nonumber\\
\langle \delta\theta_i(\tau)^2\rangle&=D\tau,\nonumber\\
\langle \theta_i(\tau)\rangle&=\left(\frac{\Omega_1+\Omega_2}{2}\right)\tau, 
\end{align}
which yield 
\begin{align}
\mathcal{Q}_i=2. 
\end{align}

\begin{figure*}[ht!]
\includegraphics[width=0.9\linewidth]{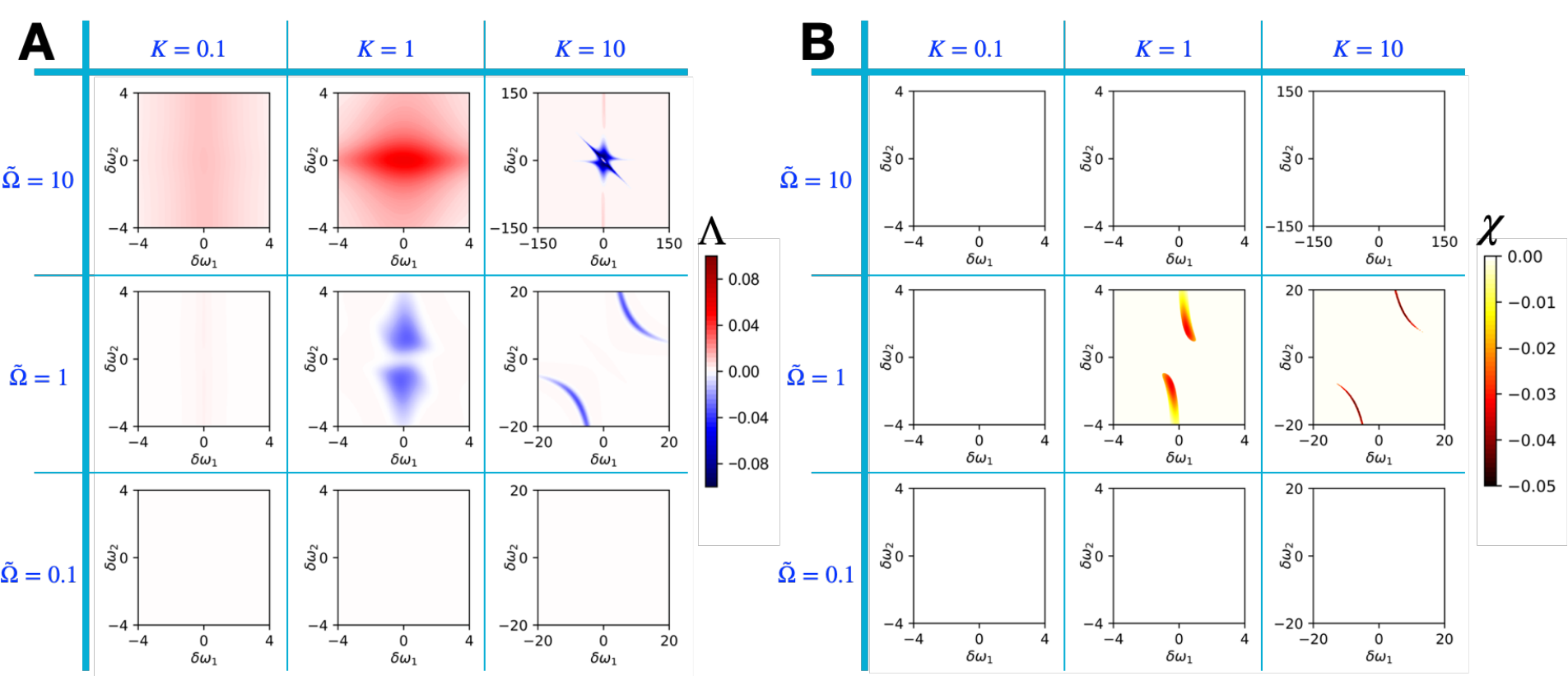}
\caption{({\bf A}) The smallest eigenvalue of the partially transposed density matrix at the steady states, $\Lambda(\delta\omega_1,\delta\omega_2)$. 
The region with negative eigenvalue (blue) characterizes the entanglement. 
({\bf B}) The overlap between $\Lambda(\delta\omega_1,\delta\omega_2)$ and $\mathcal{Q}(\delta\omega_1,\delta\omega_2)$ quantified by computing $\chi=\Theta(2-\mathcal{Q})(2/\mathcal{Q})\Lambda$, where $\Theta(\ldots)$ is the Heaviside step function. 
The region with $\chi<0$ is formed along the resonance condition. 
\label{fig:PPT} 
}
\end{figure*}

To draw an analogy with the TQS where only the 1st qubit is irradiated by the external field, 
we set $\Omega_1$ finite and $\Omega_2\rightarrow 0$ for the two coupled oscillators.  
Two limiting cases of the oscillators and their comparisons with the TQS illuminate the fundamental differences between the couplings in the TQS and in the classical oscillators. 
(i) For $K_{os}\rightarrow 0$, the uncertainty product of the 1st oscillator is minimized to the TUR bound, whereas the 2nd oscillator with $\Omega_2\rightarrow 0$ is merely subject to the thermal noise, so that we get $\mathcal{Q}_1=2$ and $\mathcal{Q}_2\gg 1$. 
This trend is also observed in the TQS when $\tilde\Omega$ is finite and $K\rightarrow 0$. 
Figure~\ref{fig:partialQ}A shows the uncertainty products of the total system ($\mathcal{Q}$) and the subsystems ($\mathcal{Q}_1$ and $\mathcal{Q}_2$) over the detunings. 
As previously discussed, 
the TQS effectively behaves as two non-interacting qubits. 
The 1st qubit subject to the external field acquires the minimal uncertainty product $\mathcal{Q}_1\approx 1.25$ at $\delta\omega_1\approx 0$. 
Meanwhile, as the 2nd qubit is effectively decoupled from the 1st qubit with no source of driving, 
its uncertainty product is large ($\mathcal{Q}_2\gg 1$, Fig.~\ref{fig:partialQ}A). 
(ii) For $K_{os}\gg |\Omega_1-\Omega_2|\approx \Omega_1$, 
the two oscillators are fully synchronized. 
Then, the current, its fluctuations, and the entropy production are reduced to half the values, yielding the uncertainty product minimized to its bound,   
$\mathcal{Q}_1=\mathcal{Q}_2=2$.  
Interestingly, in the corresponding regime for TQS ($K\gg 1$), Fig~\ref{fig:partialQ}B indicates that 
the uncertainty product of each qubit is significantly large ($\mathcal{Q}_{1,2}\gg1$) particularly under off-resonance condition.
Even under the resonance condition, each qubit's uncertainty product always exceeds 2, while the uncertainty product for the total system is still around the  minimal bound ($\mathcal{Q}\approx 1.25$) (see also Fig.~S4 and Fig.~\ref{fig:current_coherence}D, which give rise to $\mathcal{Q}$ and $\mathcal{Q}_{1,2}$).
Note that the \emph{correlated coherence}, particularly the superposition of $\ket{01}$ and $\ket{10}$ basis states (${\rm Im}[\rho_{10,01}]$, see Fig.~\ref{fig:current_coherence}C) contributes to the current from each qubit (Eqs.~\eqref{eqn:j1} and ~\eqref{eqn:j2}), while the total current originates only from the coherences of the 1st qubit (Eq.~\eqref{eqn:jtot}). 
This is a feature unique to the TQS, but lacking in the two coupled oscillators.

\subsection{Effect of entanglement on the TUR bound}
Together with the quantum coherence and quantum coupling, 
the quantum entanglement between multiple qubits, which arises in a system of multiple quantum states with inseparable correlation, can be considered another key manifestation of quantumness. 
Although several studies have explored the entanglement-powered quantum engines~\cite{wang2009thermal,park2013heat,wang2022experimental,koch2023quantum,zhang2024energy}, relatively fewer have examined the effect of entanglement on the TUR bound~\cite{prech2023entanglement}.  

Figure~\ref{fig:Q_detuning}B suggests that the strong coupling plays a key role in loosening the TUR bound; however, the coupling strength $K$ itself does not automatically imply the entanglement between the two qubits. 
To explore the effect of entanglement on the TUR bound, we apply the PPT criterion to the density matrix at steady states, $\rho^{\rm ss}$ (see Methods) and quantify the extent of entanglement present in the TQS. 

First, for the case of weak coupling $K=0.1$, it is already expected that the entanglement is absent in the TQS. Furthermore, for weak driving ($\tilde{\Omega}=0.1$), the extent of entanglement calculated based on the PPT criterion is nearly zero over $(\delta\omega_1,\delta\omega_2)$ regardless of the $K$ value. 

Any meaningful signature of entanglement is found for $K\geq 1$ and $\tilde{\Omega}\geq 1$ (Fig.~\ref{fig:PPT}). 
The parameter space giving rise to a strong entanglement does not necessarily coincide with the region with small $\mathcal{Q}$ (compare Fig.~\ref{fig:Q_detuning}A and Fig.~\ref{fig:PPT}). 
As vividly shown by the map of the smallest eigenvalue of the partially transposed density matrix ($\Lambda(\delta\omega_1,\delta\omega_2)$),  
the entanglement between the two qubits characterized with $\Lambda<0$ is the most apparent at $\tilde{\Omega}=10$ and $K=10$; yet, it contributes little to loosening the TUR bound{\color{black}, a finding consistent with a recent study on a coherent mesoscopic transport along a double quantum dot coupled to two thermal reservoirs~\cite{prech2023entanglement}.}
The overlap between the two maps simultaneously satisfying 
$\mathcal{Q}(\delta\omega_1,\delta\omega_2)\lesssim 2$ and $\Lambda(\delta\omega_1,\delta\omega_2)<0$, quantified using the overlap parameter $\chi=\Theta(2-\mathcal{Q})(2/\mathcal{Q})\Lambda$, is shaped along the resonance condition ($\delta\omega_1\delta\omega_2=K^2$) only for $\tilde{\Omega}=1$ and $K=1$, 10 (Fig.~\ref{fig:PPT}B). 

{\color{black}
\subsection{Efficiency of a stochastic engine made of TQS}
Given the quantumness characterizing the TQS, which lowers the TUR bound, 
it is of great interest to consider using TQS as a stochastic engine to convert available energy into useful work.  
With an assumption that the laser driving power $P$ is transmitted to the TQS via excitation,  
the maximum amount of power that can be extracted from TQS is $\dot{w}_{\rm max}=\hbar\omega\langle j\rangle$, which defines the efficiency, $\eta$, of the TQS as an engine
\begin{align}
\eta
=\frac{\dot{w}_{\rm max}}{P}.
\end{align}
Since the laser power is consumed to perform work $\dot{w}_{\rm max}$, 
the heat current from the system to the surrounding bath is given as $T\Sigma^{ss}=P-\hbar\omega\langle j\rangle$. 
From the inequality of the uncertainty product that specifies the TUR bound, the lower bound of non-useful heat current is written as~\cite{pietzonka2016JSM}
\begin{align}
T\Sigma^{\rm ss}\geq \mathcal{Q}_{\rm min}k_BT\langle j\rangle^2/\text{Var}(j). 
\end{align}
Thus, it follows that the efficiency for the TQS is upper-bounded as  
\begin{align}
\eta\leq\eta_{\rm TUR}&=\frac{1}{\mathcal{Q}_{\rm min}k_BT\left(\frac{\langle j\rangle}{\hbar\omega\text{Var}(j)}\right)+1}=\frac{1}{\frac{\mathcal{Q}_{\rm min}}{\mathcal{Q}}+1},
\end{align}
where the relation $\mathcal{Q}=\mathcal{A}\mathcal{F}$ in Eq.~\eqref{eqn:uncertainty_product} with $\mathcal{A}=\Delta S_{\rm tot}/k_B\langle n\rangle=\hbar\omega/k_BT$ and $\mathcal{F}=\text{Var}(j)/\langle j\rangle$ is used to obtain the last equality. 
The expression for the maximum efficiency set by the TUR bound ($\mathcal{Q}_{\rm min}$) suggests that when the TQS is employed as an engine to extract work, its efficiency is upper-bounded by $\eta_{\rm TUR}$. 
For TQS, the TUR bound ($\mathcal{Q}_{\rm min}^{\rm TQS}$) is smaller than the bound of classical counterpart, i.e., $\mathcal{Q}_{\rm min}^{\rm TQS}<2$, suggesting that the quantumness present in the TQS can enhance the engine efficiency.

\subsection{Effect of quantum nature of external driving field}
Although the current study adopts the semi-classical description of light source, 
from the quantum-mechanical perspective of light-matter interaction, it is still of interest to explore the effect of quantized light, 
particularly, on the TUR bound. In the regimes involving relatively strong external field, with Rabi frequency $\Omega$, while satisfying the requirement of RWA ($\omega\gg \Omega$), the field irradiating the TQS can be represented as a coherent state and its interaction with quantum objects is described using Jaynes-Cummings Hamiltonian, 
\begin{align}
H_{\rm JC,int} = \hbar g(\hat{a}\sigma_++\hat{a}^\dagger\sigma_-),
\end{align}
where $g$ is a coupling strength, and $\hat{a}$ and $\hat{a}^\dagger$ are the creation and annihilation operators of the quantized light field.
Here, a coherent state of light is described as 
$\ket{\alpha}=e^{-|\alpha|^2/2} \sum_{n=0}^\infty\frac{\alpha^n}{\sqrt{n!}}\ket{n}$ expanded in occupation number representation,
satisfying the relation
$\hat{a}\ket{\alpha}=\alpha\ket{\alpha}$.
We note that
in the limit of strong field, i.e., $|\alpha|^2\gg 1$, coherent states also approximately satisfy the relation, $\bra{\alpha} e^{- (i/\hbar) H_{\rm JC,int} t } \ket{\alpha} \approx e^{- i g (\alpha \sigma_+ + \alpha^* \sigma_-) t} $, which reduces to the semi-classical description of the light-matter interaction in Eq.~\eqref{eq:H_ext}.
Under this condition, our semi-classical interaction Hamiltonian remians valid,
and the effect of quantized fields is only marginal. 
In the weak field regime, on the other hand, 
higher-order corrections are no longer insignificant, and are likely contribute to the nonequilibrium current and current fluctuation. 

In addition, non-classical states of light, such as squeezed states, could significantly influence both quantum coherence and consequently current.
This influence has a potential 
to lower the TUR bound and, in turn, increase the engine efficiency beyond the classical limit~\cite{rossnagel2014nanoscale,klaers2017squeezed,niedenzu2018quantum}.

\begin{figure*}[ht!]
\includegraphics[width=0.7\linewidth]{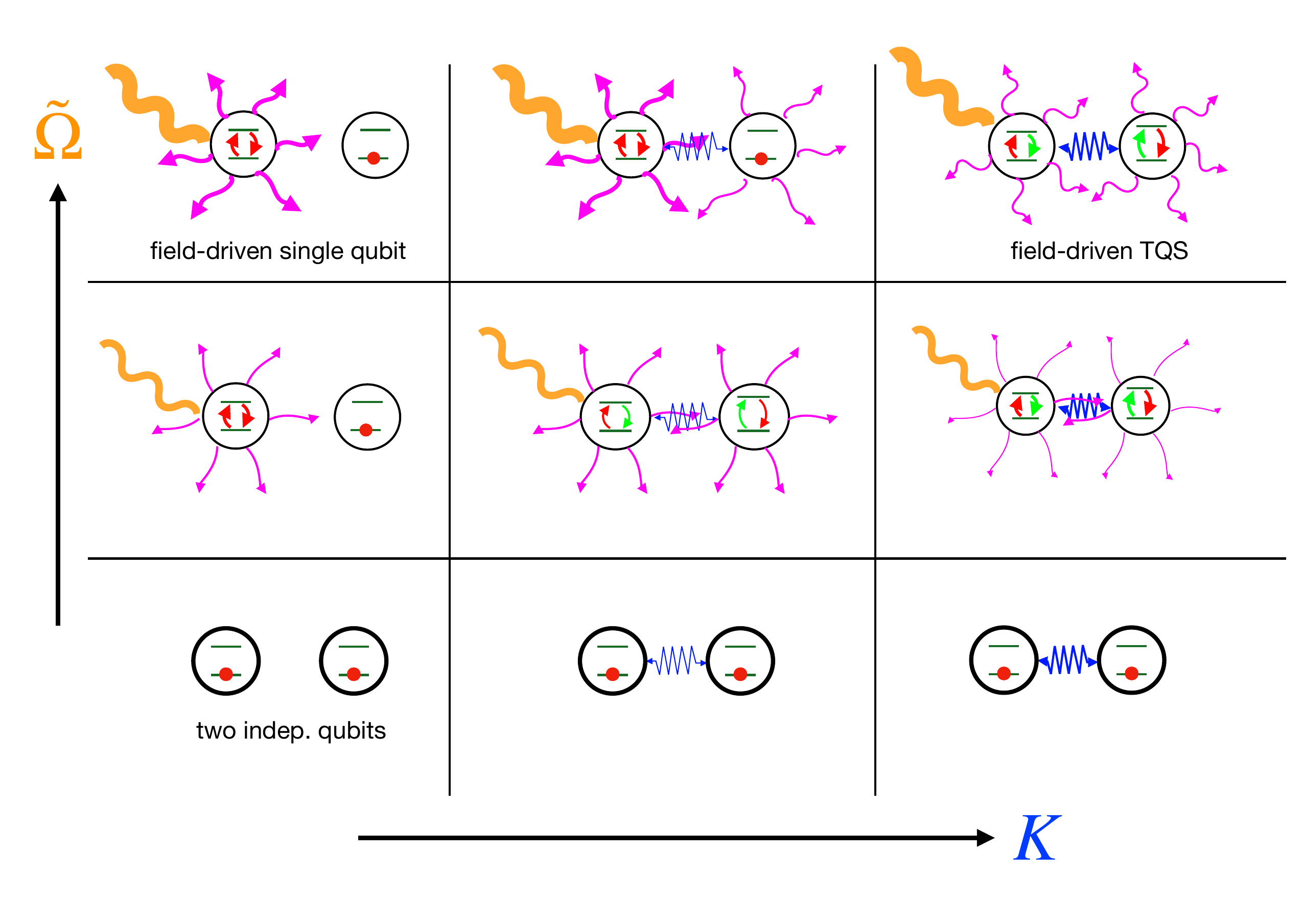}
\caption{Cartoon illustrating the dynamics of field-driven TQS with varying strength of Rabi frequency 
($\tilde{\Omega}$) and coupling ($K$).  The photon injected from the external field is depicted in orange. The dissipation from the qubit(s) is in magenta. The coupling between the two qubits are depicted with wavy arrow in blue. 
The thickness of the arrows represents the strength of the field (orange), the current (magenta) and the coupling (blue). 
The waviness of the magenta line represents the Fano factor of the photon current, with greater waviness indicating a higher Fano factor.
The extent of entanglement is represented using the red and green arrows of transition. 
\label{fig:diagram} 
}
\end{figure*}

\section{Summary}
The findings from our study on the dissipative TQS subject to an external field are summarized using the illustration in Fig.~\ref{fig:diagram}. 
\begin{enumerate}
\item When the field is weak ($\tilde{\Omega}\ll 1$), the current or current fluctuations generated from the two qubits 
is nearly zero, regardless of the coupling strength ($K$).   
In this case, $\mathcal{Q}^{\rm TQS}=\mathcal{A}\coth{\left(\mathcal{A}/2\right)}$, which converges to 2 when $\mathcal{A}\rightarrow 0$. 
\item For weak coupling ($K\ll 1$), the two qubits are effectively decoupled. 
The photons irradiated on the 1st qubit induce excitation and coherence,
followed by dissipation into the heat bath. 
The net photon current flowing through the 1st qubit, its fluctuations, and the associated entropy production are 
dictated by the TUR of the field-driven single TLS (Eq.~\eqref{eqn:Q_TLS}).
\item Provided that the external field is strong enough ($\tilde{\Omega}\geq 1$), a strong coupling ($K\gg 1$) between the two qubits is essential for creating an entanglement in the TQS, and for transmitting excitation and coherence from
the 1st qubit to the 2nd qubit (compare Fig.~\ref{fig:current_coherence}A, B with Fig.~S1A, B). 
{\color{black}The 1st and 2nd qubits share the current generation with increasing coupling strength, while the overall photon current from the system is attenuated (Eqs.~\eqref{eqn:jtot}--~\eqref{eqn:j2} and Fig.~\ref{fig:current_coherence}D).}
\item The most effective suppression of photon current fluctuations, which gives rise to TUR violation, is achieved when $\tilde{\Omega}$ is moderate, and {\color{black}the system satisfies the proper detuning condition} ($\delta\omega_1\delta\omega_2=K^2$), {\color{black}the latter of which underscores the aspect of light-matter interaction of the current study and distinguishes it from the recent two-qubit model considered in the context of a temperature gradient-driven electron-transport against chemical gradient~\cite{prech2023entanglement}.}
\item 
The TUR bound of TQS depends on the coupling strength $K$ and the field strength $\tilde{\Omega}$. 
When TQS is strongly coupled ($K\gg 1$) as well as subject to a relatively large field ($\tilde{\Omega}>3$),  its uncertainty product is 
minimized to a value smaller than the bound of classical TUR ($\mathcal{Q}_{\rm min}=2$),
\begin{align}
\mathcal{Q}^{\rm TQS}_{K\gg 1,\tilde{\Omega}}\geq\mathcal{Q}_{K\gg 1,\tilde{\Omega}>3, {\rm min}}^{\rm TQS}\approx 1.36.
\end{align}
On the other hand, when the two qubits are effectively decoupled at $K\lesssim 1$
the uncertainty product of TQS is minimized to $\mathcal{Q}_{\rm min}^{\rm TLS}\approx 1.25$,  
\begin{align}
\mathcal{Q}^{\rm TQS}_{K\lesssim 1}\geq \mathcal{Q}_{K\lesssim 1,{\rm min}}^{\rm TQS}\geq \mathcal{Q}_{\rm min}^{\rm TLS}\approx 1.25.
\end{align} 
Taken together, the TUR bound of TQS cannot be smaller than that of TLS. 
\begin{align}
\mathcal{Q}_{\rm min}^{\rm TQS}\geq \mathcal{Q}_{\rm min}^{\rm TLS}.
\label{eqn:TLS_bound} 
\end{align}
\end{enumerate}

Here are our final remarks {\color{black}with prospect.} 
The suboptimal bound of the uncertainty product, $\mathcal{Q}_{K\gg 1,\tilde{\Omega}>3, {\rm min}}^{\rm TQS}\approx 1.36$, {\color{black}which is acquired 
for strongly coupled qubits subject to a strong field (Fig.~\ref{fig:okmin}),} is specific to the TQS.  
{\color{black}Nevertheless, together with our finding from Fig.~\ref{fig:PPT} that the entanglement between two qubits plays little role in loosening the TUR bound, the inequality of Eq.~\eqref{eqn:TLS_bound} obtained based on TQS may be general enough to be extended to multi-qubit systems.}   
{\color{black}We surmise that the TLS sets the fundamental limit of TUR even for multi-qubit systems.} 
{\color{black}As an extension of TQS, 
the $N$-qubit system can be related to photosynthetic systems, more specifically light-harvesting complexes (LHC) comprised of quantum-mechanically coupled multiple chromophores irradiated by the light source~\cite{jang2018delocalized,wang2019quantum,Cogdell_Gall_Kohler_2006,Jianshu2020}. 
The problem of how the TUR of mutually or nearest-neighbor coupled qubits changes would be a topic of great interest for future investigation. }

\section{Methods}
\subsection{Characterization of entanglement}
The criterion of positive partial transpose (PPT) is employed to determine 
whether a given density matrix is separable or entangled~\cite{peres1996separability}. 
According to the criterion, the density matrix of a bipartite system is separable if all the eigenvalues of the partial-transposed density matrix are non-negative. Although this is only a necessary condition for general bipartite systems to be separable, the criterion serves as both the sufficient and necessary condition for the separability of two-qubit systems~\cite{Horodecki1996PLA}. If the partial transpose of a two-qubit density matrix has any negative eigenvalues, the two qubits are no longer separable to any independent basis, thus being entangled. 

The PPT criterion applied to our TQS amounts to calculating the eigenvalue of the density matrix where the 1st qubit is transposed, i.e.,  
\begin{align}
(\rho^{\rm ss})^{\mathcal{T}_1}=
\begin{pmatrix}
\rho_{11,11}^{\rm ss}&\rho_{11,10}^{\rm ss}&\rho_{01,11}^{\rm ss}&\rho_{01,10}^{\rm ss}\\
\rho_{10,11}^{\rm ss}&\rho_{10,10}^{\rm ss}&\rho_{00,11}^{\rm ss}&\rho_{00,10}\\
\rho_{11,01}^{\rm ss}&\rho_{11,00}^{\rm ss}&\rho_{01,01}^{\rm ss}&\rho_{01,00}^{\rm ss}\\
\rho_{10,01}^{\rm ss}&\rho_{10,00}^{\rm ss}&\rho_{00,01}^{\rm ss}&\rho_{00,00}^{\rm ss}
\end{pmatrix}. 
\end{align}
If the smallest eigenvalue of the partially transposed matrix, $(\rho^{\rm ss})^{\mathcal{T}_1}$, which is defined as $\Lambda$ (Fig.~\ref{fig:PPT}), is more negative, 
the entanglement is deemed stronger. 
 
\subsection{Method of cumulant generating function}
Our objective in this section is to introduce a systematic means to calculate 
the mean steady-state current 
$\langle j \rangle$ and the variance ${\rm Var}(j)$ from the reduced density matrix. 
We first formally define the cumulant generating function:  
\begin{align}
\mathcal{G}(z,\tau)=\ln{\langle e^{zn}\rangle}=\ln{\sum_{n=-\infty}^{\infty}P(n,\tau)e^{zn}},
\end{align} 
where $P(n,\tau)$ is the probability of $n$ net photons being emitted from the TQS during the time interval $\tau$, which is the sum of corresponding four populations that can similarly be defined using the reduced density matrix $\rho(n,\tau)$, namely, $P(n,\tau)=\rho_{11,11}(n,\tau)+\rho_{10,10}(n,\tau)+\rho_{01,01}(n,\tau)+\rho_{00,00}(n,\tau)={\rm Tr}\left(\rho(n,\tau)\right)$ with $P(n,\tau)$ satisfying the normalization condition, $\sum\limits_{n=-\infty}^{\infty}P(n,\tau)=1$. 
Thus, for a given $\mathcal{G}(z,\tau)$, one can calculate the $k$-th cumulant using 
\begin{align}
\langle\langle n^k\rangle\rangle(\tau)=\frac{\partial^k\mathcal{G}(z,\tau)}{\partial z^k}\Big|_{z=0}. 
\label{eqn:cumulant}
\end{align}

Next, the vectorized form of density matrix defined in the Fock-Liouville space, $\tilde{\rho}$, can be decomposed into $\tilde{\rho}(n,\tau)$ of all possible $n$ as 
\begin{align}
\tilde{\rho}(\tau)=\sum_{n=-\infty}^\infty\tilde{\rho}(n,\tau). 
\end{align}
Here, $\tilde{\rho}(n,\tau)$ is expected to satisfy the $n$-resolved Master equation: 
\begin{align}
    \partial_t\tilde{\rho}(n,t)=\mathcal{L}_0\tilde{\rho}(n,t)
    +\underbrace{\mathcal{L}_+\tilde{\rho}(n-1,t)}_{\text{emission}}
    +\underbrace{\mathcal{L}_-\tilde{\rho}(n+1,t)}_{\text{absorption}}, 
    \label{eqn:n_resolved}
\end{align}
where 
we have introduced the total emission and absorption operators from the two qubits, $\mathcal{L}_\pm=\mathcal{L}^1_\pm +\mathcal{L}^2_\pm$ and defined the remaining elements as $\mathcal{L}_0\equiv \mathcal{L}-\mathcal{L}_+-\mathcal{L}_-$. 
The two last terms on the right hand side of Eq.~\eqref{eqn:n_resolved} 
express the contributions of emission and absorption to the evolution of $\tilde{\rho}(n,t)$ with $\mathcal{L}_\pm$. 
For the TQS, these operators can be written explicitly as:
\begin{align}
    \mathcal{L}^1_+ &=\mathcal{L}_{0000,1010}
    +\mathcal{L}_{0001,1011}
    +\mathcal{L}_{0100,1110}
    +\mathcal{L}_{0101,1111}\nonumber\\
    \mathcal{L}^1_-&=\mathcal{L}_{1010,0000}
    +\mathcal{L}_{1011,0001}
    +\mathcal{L}_{1110,0100}
    +\mathcal{L}_{1111,0101}\nonumber\\
    \mathcal{L}^2_+&=\mathcal{L}_{0000,0101}
    +\mathcal{L}_{0010,0111}
    +\mathcal{L}_{1000,1101}
    +\mathcal{L}_{1010,1111}\nonumber\\
    \mathcal{L}^2_-&=\mathcal{L}_{0101,0000}
    +\mathcal{L}_{0111,0010}
    +\mathcal{L}_{1101,1000}
    +\mathcal{L}_{1111,1010}
    \label{eqn:operator_individual}.
\end{align}
Specifically, $\mathcal{L}_{0001,1011}$ is the one of the elements of the Liouville operator associated with the transition: $\rho_{10,11}\rightarrow \rho_{00,01}$. 

Further considering a discrete Laplace transform of $\tilde{\rho}(n,\tau)$, $\hat{\rho}_z(\tau)\equiv \sum\limits_{n=-\infty}^\infty\tilde{\rho}(n,\tau)e^{zn}$, 
we can recast the $n$-resolved Master equation in Eq.~\eqref{eqn:n_resolved} to a more simplifying form of the linear differential equation
\begin{align}
    \partial_\tau \hat{\rho}_z (\tau) =\mathcal{L}(z)\hat{\rho}_z (\tau)
\label{eqn:recast}
\end{align}
with a modified $16\times 16$ Liouville operator defined as 
\begin{align}
\mathcal{L}(z)=\mathcal{L}_0+e^z \mathcal{L}_++e^{-z} \mathcal{L}_-.
\label{eqn:mod_Liouville_operator}
\end{align} 
A formal solution of Eq.~\eqref{eqn:recast} is written as 
\begin{align}
\hat{\rho}_z(\tau)\equiv \sum\limits_{n=-\infty}^\infty\tilde{\rho}(n,\tau)e^{zn}=e^{\tau\mathcal{L}(z)}\hat{\rho}_z(0)=\sum_k\vec{l}_ke^{\lambda_k(z)\tau},
\end{align}
where $0>\lambda_0(z)>\lambda_1(z)>\cdots$ is satisfied, and 
for $\tau\gg 1$ the solution is dominated by the largest eigenvalue $\lambda_0(z)$, 
and 
it is expected that $\lambda_0(0)=0$ and $\vec{l}_0=\tilde{\rho}^{\rm ss}$. 
Therefore, 
\begin{align}
\ln{\left(\sum_{n=-\infty}^\infty P(n,\tau)e^{zn}\right)}\sim\lambda_0(z)\tau 
\end{align}
should hold for $\tau\gg 1$. 
Thus, at steady states ($\tau\gg 1$), the cumulant generating function expressed in terms of the largest eigenvalue of the modified Liouville operator
$\mathcal{G}(z,\tau)
\sim \lambda_0(z)\tau$ along with Eq.~\eqref{eqn:cumulant}  
allows us to relate the $k$-th cumulant with the largest eigenvalue of $\mathcal{L}(z)$, i.e., $\lambda_0(z)$ as follows:
\begin{align}
\lim_{\tau\gg 1}\frac{\langle\langle n^k\rangle\rangle(\tau)}{\tau}=\frac{\partial^k\lambda_0(z)}{\partial z^k}\Big|_{z=0}. 
\label{eqn:cumulant_eig}
\end{align}

Instead of directly solving the eigenvalue problem via the characteristic polynomial, 
\begin{align} 
p_{\mathcal{L}}(\lambda)
\equiv\det\left(\lambda(z)\mathcal{I}-\mathcal{L}(z)\right)=
\sum_{i=0}^{16} a_i(z)\lambda^i(z)=0,  
\label{eq:characteristic_poly}
\end{align}
we differentiate both sides of Eq.~\eqref{eq:characteristic_poly} with respect to  $z$ and taking $z=0$ and noting that $\lambda_0(0)=0$, 
we obtain
\begin{align}
a'_0(0)+a_1(0)\lambda'_0(0)=0. 
\end{align}
Together with Eq.~\eqref{eqn:cumulant_eig} with $k=1$, the photon current is related to the coefficients of the characteristic polynomial as follows. 
\begin{align}
\langle j \rangle = \lim_{t\rightarrow\infty}\frac{\langle n\rangle(t)}{t}= \lambda_0'(0)=-\frac{a'_0(0)}{a_1(0)}.
\label{eqn:current_expr}
\end{align}

Alternatively, by using the Cayley-Hamilton theorem we obtain the following matrix identity:
\begin{align}
p_\mathcal{L}&(\mathcal{L})=\sum_{i=0}^{16} a_i(z)\mathcal{L}^i(z)
={\bf O}_{16\times 16},
\label{eq:Cayley_Hamilton_formula}
\end{align}  
where ${\bf O}_{l \times m}$ is an $l \times m$ matrix with zero entries. We then note that $\mathcal{L}(0) \tilde\rho^{\rm ss}=\mathbf{0}_{16 \times 1}$ from the steady state condition and $\sum_{p,q} {\cal L}_{pqpq,abcd} = 0$ for all $a,b,c,d$ from the trace-preserving condition. These conditions lead to
\begin{align} 
0&= \frac{\partial p_{\cal L}({\cal L})}{\partial z}\Big|_{z=0} \tilde\rho^{\rm ss} \nonumber\\
& =\sum_{p,q}\left[ \sum_{i=0}^{16} \left[ a'_i(z) {\cal L}^i(z) + a_i(z) \partial_z {\cal L}^i(z) \right]\Big|_{z=0}\tilde\rho^{\rm ss}\right]_{pq,pq}\nonumber\\
&=a'_0(0) + a_1(0) \sum_{p,q} \left[ {\cal L}'(0)\tilde\rho^{\rm ss} \right]_{pq,pq},
\end{align} 
where we used $\sum_{p,q} \tilde\rho^{\rm ss}_{pq,pq} = {\rm Tr}[\rho^{\rm ss}] =1$.
Together with Eq.~\eqref{eqn:current_expr} and $\mathcal{L}'(0)=\mathcal{L}_+-\mathcal{L}_-$ from Eq.~\eqref{eqn:mod_Liouville_operator},  
we can derive a physically motivating, general form of the current, 
\begin{align}
\langle j \rangle = \sum_{p,q} \left[ ({\cal L}_+ - {\cal L}_- )\tilde\rho^{\rm ss} \right]_{pq,pq}, 
\label{eqn:current_general}
\end{align}
which can be used to derive the expressions in Eqs.~\eqref{eqn:j1_explicit} and ~\eqref{eqn:j2_explicit} together with Eq.~\eqref{eqn:operator_individual}. 


To obtain the current from each qubit, 
the total absorption and emission operators can be redefined as $\mathcal{L}_\pm = \mathcal{L}^1_\pm$ or $\mathcal{L}^2_\pm$  instead of summing up the entire contribution from the two qubits as  $\mathcal{L}_\pm = \mathcal{L}^1_\pm+\mathcal{L}^2_\pm$. 
The expressions for the Liouville operators associated with emissions 
($\mathcal{L}_+^1$, $\mathcal{L}_+^2$) and 
absorptions ($\mathcal{L}_-^1$, $\mathcal{L}_-^2$) given in Eq.~\eqref{eqn:operator_individual} 
offer the expressions in Eqs.~\eqref{eqn:j1_explicit} and~\eqref{eqn:j2_explicit}. 

We can continue on the similar procedure 
to obtain the current fluctuations, ${\rm Var}(j)$. 
First, from Eq.~\eqref{eqn:cumulant_eig} with $k=2$ and the second derivative of Eq.~\eqref{eq:characteristic_poly}, we get  
\begin{widetext}
\begin{align}
{\rm Var}(j)&=\lim_{t\rightarrow\infty}\frac{\langle\langle n^2\rangle\rangle(t)}{t}=\lambda_0''(0)=-\frac{a''_0(0)+2a^\prime_1(0)\langle j\rangle+2a_2(0)\langle j\rangle^2}{a_1(0)}. 
\label{eqn:var_expr}
\end{align}
Next, the second derivative of Eq.\eqref{eq:Cayley_Hamilton_formula} at $z=0$ that  operates on $\tilde{\rho}^{\rm ss}$ gives  
\begin{align} 
0&=\frac{\partial^2 p_{\cal L}({\cal L})}{\partial z^2}\Big|_{z=0} \tilde\rho^{\rm ss} =\sum_{p,q}\left[ \sum_{i=0}^{16} \left[ a''_i(z) {\cal L}^i(z) + 2 a'_i(z) \partial_z {\cal L}^i(z) + a_i(z) \partial^2_z {\cal L}^i(z) \right]\Big|_{z=0}\tilde\rho^{\rm ss}\right]_{pq,pq}\nonumber\\
&=a''_0(0) + 2 a'_1(0) \underbrace{\sum_{p,q} \left[ {\cal L}'(0)\tilde\rho^{\rm ss} \right]_{pq,pq}}_{=\langle j\rangle} + a_1(0) \underbrace{\sum_{p,q} \left[ {\cal L}''(0)\tilde\rho^{\rm ss} \right]_{pq,pq}}_{=\sum\limits_{p,q}[({\cal L}_++{\cal L}_-)\tilde{\rho}^{\rm ss}]_{pq,pq}} + \sum_{i \geq 2} 2a_i(0) \sum_{p,q} \left[ {\cal L}'(0) {\cal L}^{i-2}(0) {\cal L}'(0)\tilde\rho^{\rm ss} \right]_{pq,pq}. 
\label{eqn:Cayley-2nd}
\end{align} 
Thus, by comparing Eq.~\eqref{eqn:Cayley-2nd} with  Eq.~\eqref{eqn:var_expr}, we obtain a formal expression for the current fluctuations. 
\begin{align}
{\rm Var}(j)
&= \sum_{p,q} \left[ ({\cal L}_+ + {\cal L}_-) \tilde\rho^{\rm ss} \right]_{pq,pq} -
\frac{2 a_2(0)}{a_1(0)}  \left[\langle j\rangle^2 - \sum_{i \geq 2} \frac{a_i(0)}{a_2(0)} \sum_{p,q} \left[ ({\cal L}_+ - {\cal L}_-) {\cal L}^{i-2}({\cal L}_+ - {\cal L}_-) \tilde\rho^{\rm ss} \right]_{pq,pq}\right].
\label{eqn:fluctuations_general}
\end{align}
\end{widetext}
By evaluating Eq.~\eqref{eqn:fluctuations_general} explicitly for TLS, 
we recover the photon current fluctuations given in Eq.~\eqref{eqn:Var_j}.

\begin{acknowledgments}
This work was supported by the KIAS Individual Grant AP100301 (K.H.C.), CG085302 (H.K.), and CG035003 (C.H.) at the Korea Institute for Advanced Study. We thank the Center for Advanced Computation in KIAS for providing computing resources.
\end{acknowledgments}

\bibliography{apssamp.bib,mybib2.bib}
\clearpage 
\setcounter{equation}{0}
\setcounter{figure}{0}

\renewcommand{\theequation}{S\arabic{equation}}
\renewcommand{\thefigure}{S\arabic{figure}} 

\begin{widetext}
\section*{Supplemental Information}

Each of the density matrix elements in Eq.~\eqref{eqn:Liouville_eq} evolves as follows. 
\begin{align}
\dot{\rho}_{11,11}&=-[\gamma(\nbar_2+1)+\left(\nbar_1+1\right)]\rho_{11,11}+\nbar_1\rho_{01,01}+\gamma\nbar_2\rho_{10,10}-i\underbrace{\left(-\Omega\rho_{01,11}+\Omega\rho_{11,01}\right)}_{=2i\Omega{\rm Im}[\rho_{11,01}]}\nonumber\\
   \dot{\rho}_{11,10}&=\nbar_1\rho_{01,00}-\left[(\nbar_1+1)+\frac{\gamma}{2}(2\nbar_2+1)\right]\rho_{11,10}-i\left(-\Omega\rho_{01,10}+\Omega\rho_{11,00}-K\rho_{11,01}+\delta\omega_2\rho_{11,10}\right)\nonumber\\
    \dot{\rho}_{11,01}&=\gamma\nbar_2\rho_{10,00}-\left[\frac{1}{2}(2\nbar_1+1)+\gamma(\nbar_2+1)\right]\rho_{11,01}-i\left(-\Omega\rho_{01,01}+\delta\omega_1\rho_{11,01}-K\rho_{11,10}+\Omega\rho_{11,11}\right)\nonumber\\
     \dot{\rho}_{11,00}&=-\left[\frac{1}{2}(2\nbar_1+1)+\frac{\gamma}{2}(2\nbar_2+1)\right]\rho_{11,00}-i\left(-\Omega\rho_{01,00}+(\delta\omega_1+\delta\omega_2)\rho_{11,00}+\Omega\rho_{11,10}\right)\nonumber\\
    \dot{\rho}_{10,11}&=\nbar_1\rho_{00,01}-\left[(\nbar_1+1)+\frac{\gamma}{2}(2\nbar_2+1)\right]\rho_{10,11}-i\left(-\Omega\rho_{00,11}+K\rho_{01,11}+\Omega\rho_{10,01}-\delta\omega_2\rho_{10,11}\right)\nonumber\\
     \dot{\rho}_{10,10}&=\nbar_1\rho_{00,00}-\left[(\nbar_1+1)+\gamma\nbar_2\right]\rho_{10,10}+\gamma(\nbar_2+1)\rho_{11,11}-i\underbrace{\left[\Omega(\rho_{10,00}-\rho_{00,10})+K(\rho_{01,10}-\rho_{10,01})\right]}_{=2i\Omega{\rm Im}[\rho_{10,00}]-2iK{\rm Im}[\rho_{10,01}]}\nonumber\\
    \dot{\rho}_{10,01}&=-\left[\frac{1}{2}(2\nbar_1+1)\rho_{10,01}+\frac{\gamma}{2}(2\nbar_2+1)\right]\rho_{10,01}-i\left(-\Omega\rho_{00,01}+K\rho_{01,01}+(\delta\omega_1-\delta\omega_2)\rho_{10,01}-K\rho_{10,10}+\Omega\rho_{10,11}\right)\nonumber\\
    \dot{\rho}_{10,00}&=-\left[\frac{1}{2}(2\nbar_1+1)+\gamma\nbar_2\right]\rho_{10,00}+\gamma(\nbar_2+1)\rho_{11,01}-i\left(-\Omega\rho_{00,00}+K\rho_{01,00}+\delta\omega_1\rho_{10,00}+\Omega\rho_{10,10}\right)\nonumber\\
        \dot{\rho}_{01,11}&=\gamma\nbar_2\rho_{00,10}-\left[\frac{1}{2}(2\nbar_1+1)+\gamma(\nbar_2+1)\right]\rho_{01,11}-i\left(\Omega\rho_{01,01}-\delta\omega_1\rho_{01,11}+K\rho_{10,11}-\Omega\rho_{11,11}\right)\nonumber\\
    \dot{\rho}_{01,10}&=-\left[\frac{1}{2}(2\nbar_1+1)+\frac{1}{2}\gamma(2\nbar_2+1)\right]\rho_{01,10}-i\left(\Omega\rho_{01,00}-K\rho_{01,01}-(\delta\omega_1-\delta\omega_2)\rho_{01,10}+K\rho_{10,10}-\Omega\rho_{11,10}\right)\nonumber\\
     \dot{\rho}_{01,01}&=\gamma\nbar_2\rho_{00,00}-[\nbar_1+\gamma(\nbar_2+1)]\rho_{01,01}+(\nbar_1+1)\rho_{11,11}-i\underbrace{\left[K(\rho_{10,01}-\rho_{01,10})-\Omega(\rho_{11,01}-\rho_{01,11})\right]}_{=2iK{\rm Im}[\rho_{10,01}]-2i\Omega{\rm Im}[\rho_{11,01}]}\nonumber\\
    \dot{\rho}_{01,00}&=-\left[\nbar_1+\frac{\gamma}{2}(2\nbar_2+1)\right]\rho_{01,00}+(\nbar_1+1)\rho_{11,10}-i\left(\delta\omega_2\rho_{01,00}+K\rho_{10,00}+\Omega\rho_{01,10}-\Omega\rho_{11,00}\right)\nonumber\\
    \dot\rho_{00,11}&=-\left[\frac{1}{2}(2\nbar_1+1)+\frac{\gamma}{2}(2\nbar_2+1)\right]\rho_{00,11}-i\left(\Omega\rho_{00,01}-(\delta\omega_1+\delta\omega_2)\rho_{00,11}-\Omega\rho_{10,11}\right)\nonumber\\
    \dot\rho_{00,10}&=-\left[\frac{1}{2}(2\nbar_1+1)+\gamma\nbar_2\right]\rho_{00,10}+\gamma(\nbar_2+1)\rho_{01,11}-i\left(\Omega\rho_{00,00}-K\rho_{00,01}-\delta\omega_1\rho_{00,10}-\Omega\rho_{10,10}\right)\nonumber\\
    \dot\rho_{00,01}&=-\left[\nbar_1+\frac{\gamma}{2}(2\nbar_2+1)\right]\rho_{00,01}+(\nbar_1+1)\rho_{10,11}-i\left(-\delta\omega_2\rho_{00,01}-K\rho_{00,10}-\Omega\rho_{10,01}+\Omega\rho_{00,11}\right)\nonumber\\
    \dot\rho_{00,00}&=-(\nbar_1+\gamma\nbar_2)\rho_{00,00}+\gamma(\nbar_2+1)\rho_{01,01}+(\nbar_1+1)\rho_{10,10}-i\underbrace{\left(\Omega\rho_{00,10}-\Omega\rho_{10,00}\right)}_{=-2i\Omega{\rm Im}[\rho_{10,00}]}
    \end{align}
\\

The steady state conditions for the population, 
$\dot{\rho}_{11,11}=0$, $\dot{\rho}_{10,10}=0$, $\dot{\rho}_{01,01}=0$, and $\dot{\rho}_{00,00}=0$, yield
\begin{align}
2\Omega{\rm Im}[\rho_{11,01}^{\rm ss}]=[(\nbar_1+1)\rho_{11,11}^{\rm ss}-\nbar_1\rho_{01,01}^{\rm ss}]+[\gamma(\nbar_2+1)\rho_{11,11}^{\rm ss}-\gamma\nbar_2\rho_{01,01}^{\rm ss}]
\label{eqn:rho1111}
\end{align}
\begin{align}
-2\Omega{\rm Im}[\rho_{10,00}^{\rm ss}]+2K{\rm Im}[\rho_{10,01}^{\rm ss}]=
-[(\nbar_1+1)\rho_{10,10}^{\rm ss}-\nbar_1\rho_{00,00}^{\rm ss}]+[\gamma(\nbar_2+1)\rho_{11,11}^{\rm ss}-\gamma\nbar_2\rho_{10,10}^{\rm ss}]
\label{eqn:rho1010}
\end{align}
\begin{align}
2\Omega{\rm Im}[\rho_{11,01}^{\rm ss}]-2K{\rm Im}[\rho_{10,01}^{\rm ss}]=
[(\nbar_1+1)\rho_{11,11}^{\rm ss}-\nbar_1\rho_{01,01}^{\rm ss}]-[\gamma(\nbar_2+1)]\rho_{01,01}^{\rm ss}-\gamma\nbar_2\rho_{00,00}^{\rm ss}]
\label{eqn:rho0101}
\end{align}
\begin{align}
2\Omega{\rm Im}[\rho_{10,00}^{\rm ss}]=
[(\nbar_1+1)\rho_{10,10}^{\rm ss}-\nbar_1\rho_{00,00}^{\rm ss}]+[\gamma(\nbar_2+1)\rho_{01,01}^{\rm ss}-\gamma\nbar_2\rho_{00,00}^{\rm ss}]
\label{eqn:rho0000}
\end{align}
Therefore, together with the current from each qubit given in Eqs.~\eqref{eqn:j1_explicit} and \eqref{eqn:j2_explicit}, 
subtracting Eq.~(\ref{eqn:rho0101}) from Eq.~(\ref{eqn:rho1111}) or adding Eq.~(\ref{eqn:rho1010}) and Eq.~(\ref{eqn:rho0000}) yields  Eq.~\eqref{eqn:j2}, whereas adding Eq.~(\ref{eqn:rho1111}) and Eq.~(\ref{eqn:rho0000}) yields Eq.~\eqref{eqn:jtot}.  

\end{widetext}

\begin{figure*}[ht!]
\includegraphics[width=0.9\linewidth]{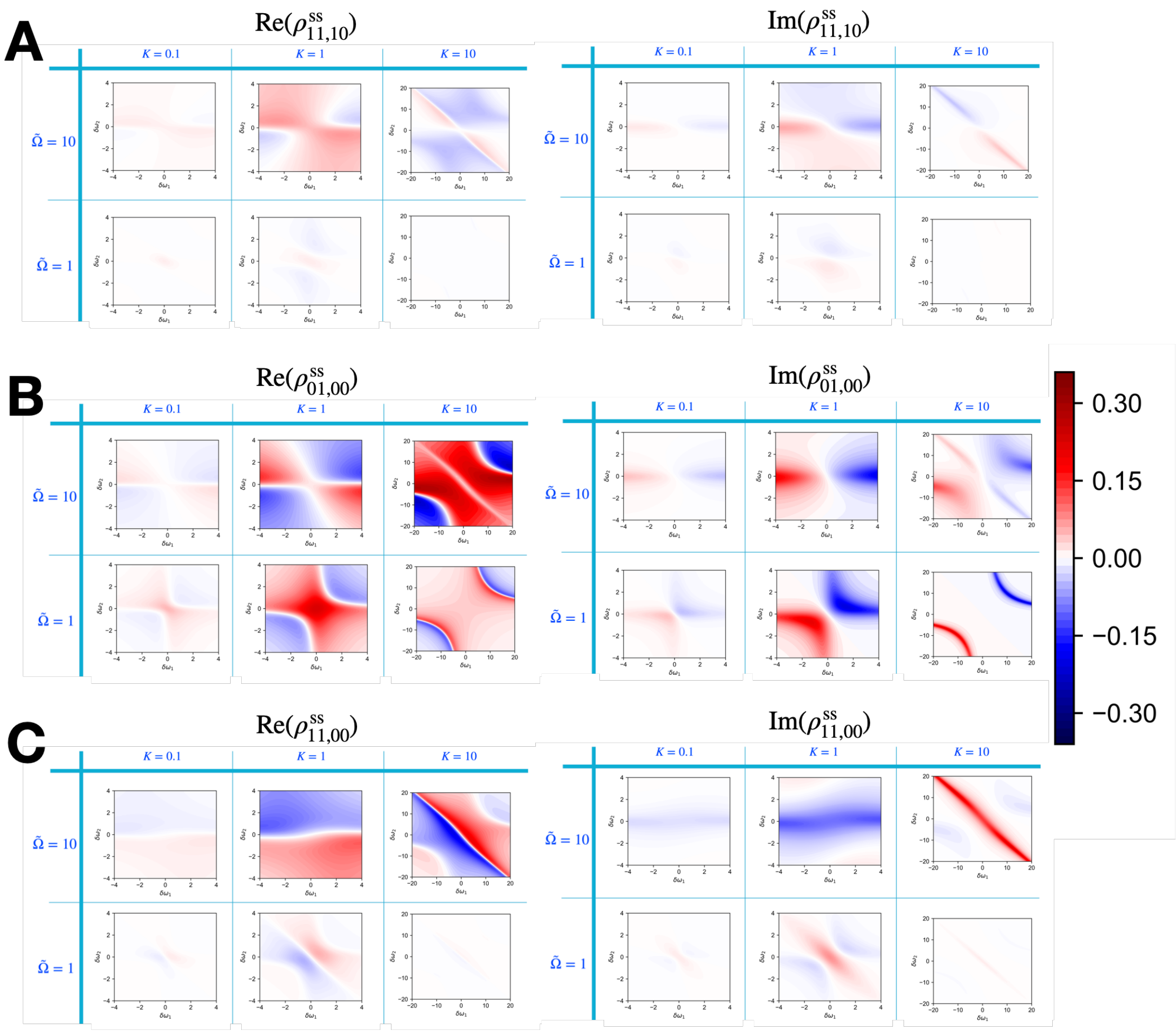}
\caption{Real and imaginary parts of coherence of the 2nd qubit when the 1st qubit is in its (A) excited and (B) ground states. 
(C) The correlated coherence between $\ket{11}$ and $\ket{00}$.
}
\label{fig:coherence_11} 
\end{figure*}


\begin{figure*}[ht!]
\includegraphics[width=0.8\linewidth]{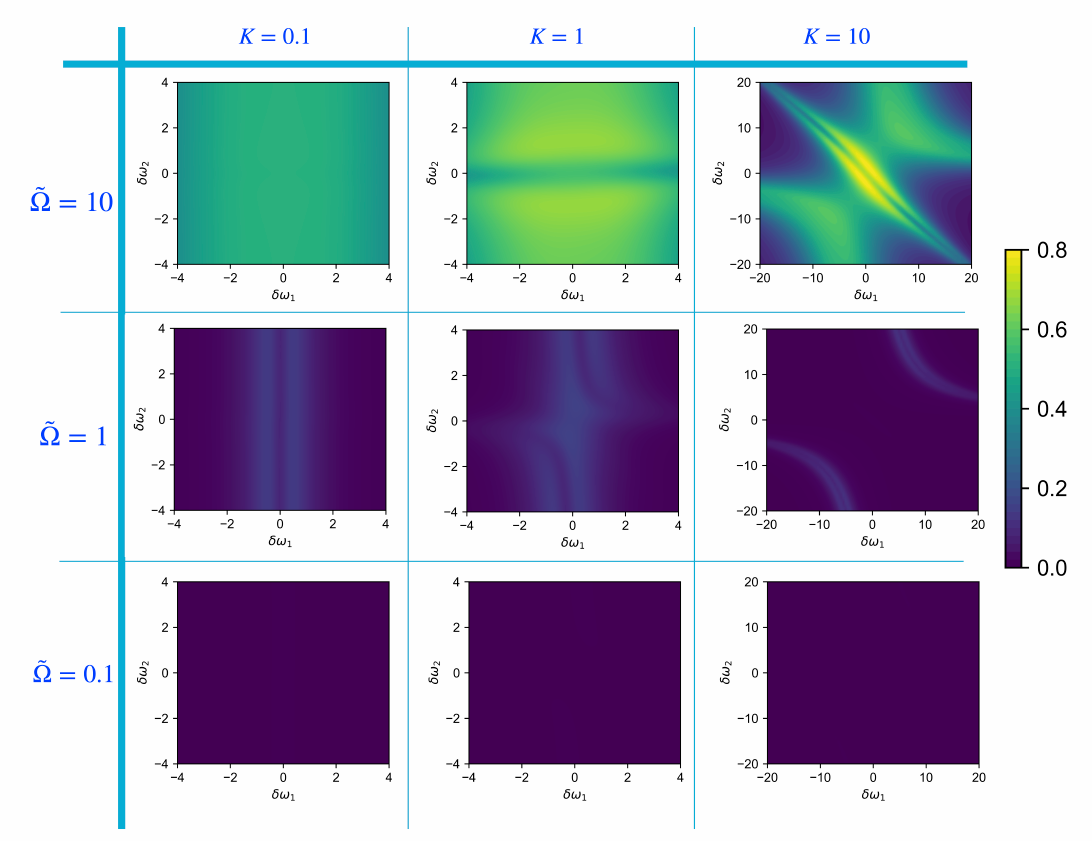}
\caption{Photon current fluctuations over varying amount of detunings calculated for 9 different sets of parameters ($K=0.1$, 1, 10 and $\tilde{\Omega}=0.1$, 1, 10). }
\label{fig:variance}
\end{figure*}


\begin{figure}[ht!]
\includegraphics[width=0.8\linewidth]{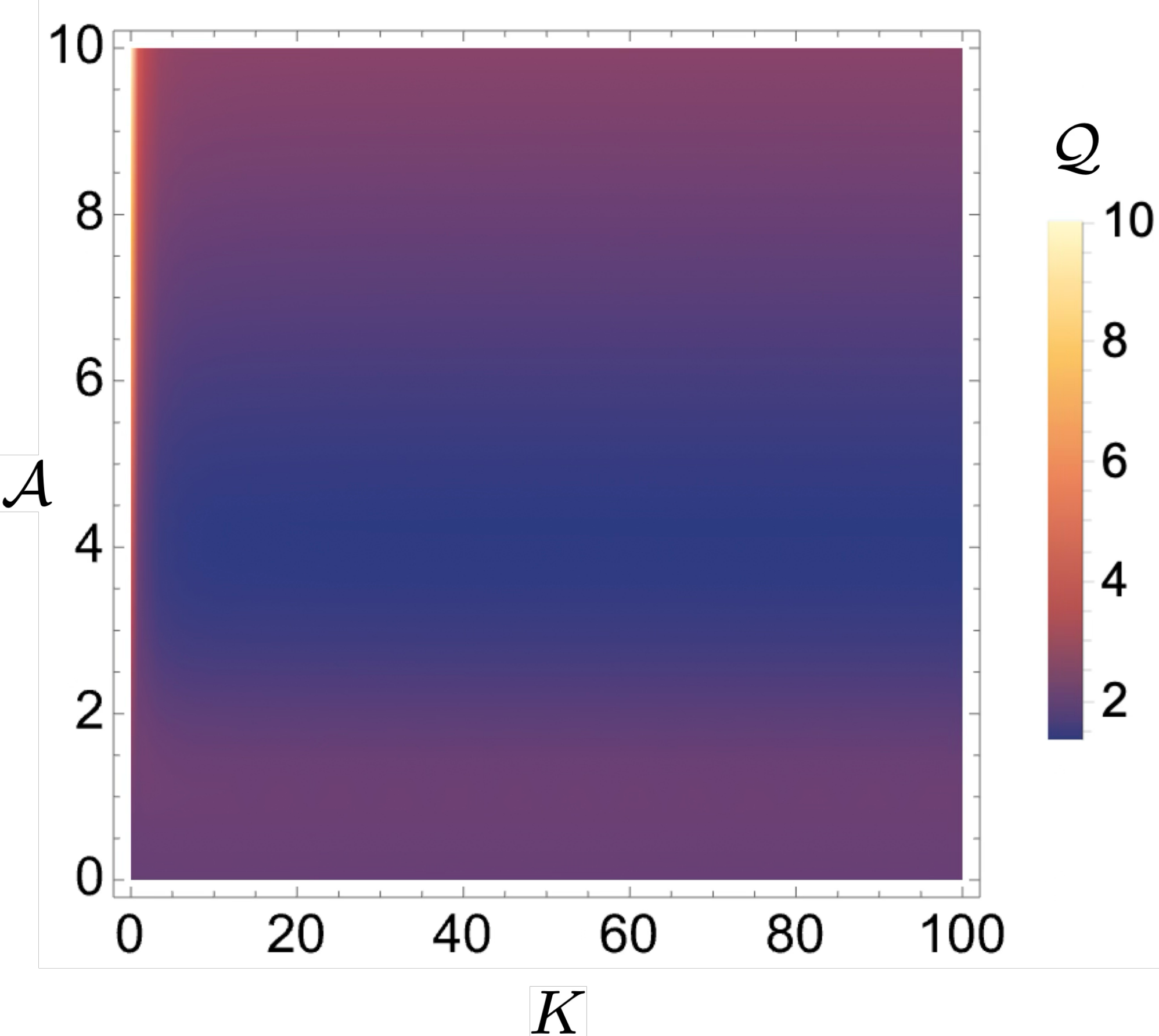}
\caption{$\mathcal{Q}$ as a function of $K$ and $\mathcal{A}$ under the constraints of 
$\delta\omega_1=K\cdot\tilde{\Omega}$ and $\delta\omega_2=K/\tilde{\Omega}$. 
We also made another constraint $\tilde\Omega$. 
For large $K(\gg 1)$, and thus large $\tilde\Omega$, the minimal $\mathcal{Q}$ is obtained when $\mathcal{A}\approx 4.08$. 
}
\label{fig:qka} 
\end{figure}

\begin{figure*}[ht!]
\includegraphics[width=0.75\linewidth]{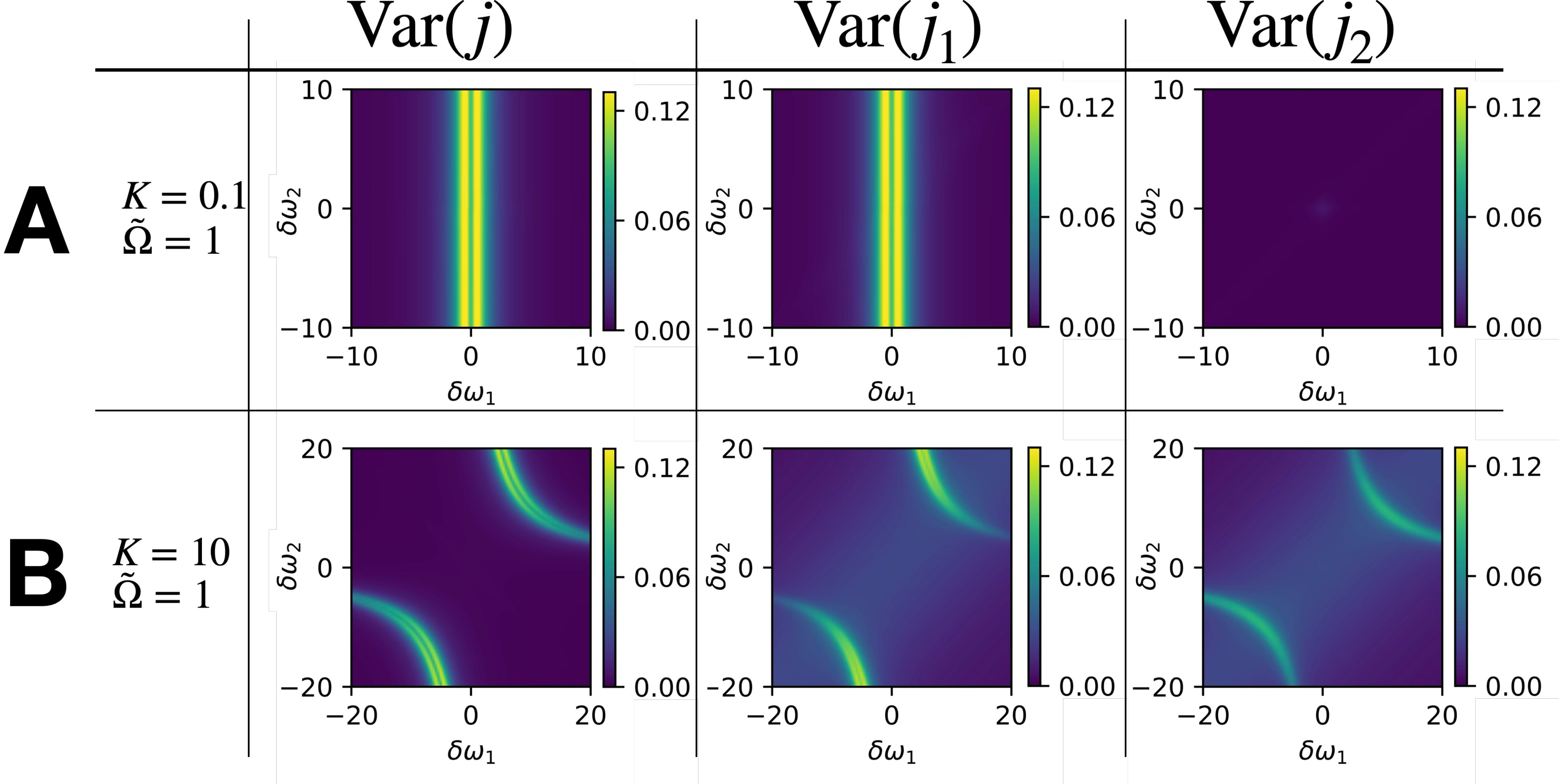}
\caption{Photon current fluctuations of entire system and each qubit as a function of detunings ($\delta\omega_1, \delta\omega_2$).
}
\label{fig:partialV} 
\end{figure*}


\end{document}